\documentclass[graybox]{svmult}

 \usepackage{mathptmx}       % selects Times Roman as basic font
\usepackage{helvet}         % selects Helvetica as sans-serif font
\usepackage{courier}        % selects Courier as typewriter font
\usepackage{type1cm}        % activate if the above 3 fonts are
\usepackage{makeidx}         % allows index generation
\usepackage{graphicx}        % standard LaTeX graphics tool
 \usepackage{subfig} 

 \usepackage[numbers] {natbib}

\usepackage{multicol}        % used for the two-column index
\usepackage{multirow}        % used for the two-column index
\usepackage[bottom]{footmisc}% places footnotes at page bottom

\makeindex             % used for the subject index
\emergencystretch=\maxdimen
\begin{document}

\title*{ On the time delay evolution of five  Active Galactic Nuclei}
\author{A. Kova{\v c}evi{\'c},  L.  {\v C}.  Popovi{\'c},  A. I. Shapovalova,  D. Ili{\'c},  A. N. Burenkov and  V. H. Chavushyan}
\institute{A. Kova{\v c}evi{\'c}, D. Ili{\'c} \at Department of astronomy, Faculty of mathematics, University of Belgrade, Studentski trg 16, 11000 Belgrade, Serbia \email{andjelka@matf.bg.ac.rs,  dilic@matf.bg.ac.rs}
\and L. {\v C}. Popovi{\'c} \at  Astronomical Observatory Belgrade, Volgina 7, 11060 Belgrade, Serbia, \email{lpopovic@aob.rs}
\and A. I. Shapovalaova, A. N. Burenkov  \at Special Astrophysical Observatory of the Russian AS, Nizhnij Arkhyz, 
Karachaevo - Cherkesia 369167, Russia,  \email{ashap@sao.ru, banbur@gmail.com}
%\and  D. Ili{\'c},\at Department of astronomy, Faculty of mathematics, University of Belgrade , Studentski trg 16, 11000 Belgrade, Serbia \email{dilic@matf.bg.ac.rs}
%\and  A. N. Burenkov,  \at Special Astrophysical Observatory of the Russian AS, Nizhnij Arkhyz, 
%Karachaevo - Cherkesia 369167, Russia \email{banbur@gmail.com}
\and V. H. Chavushyan \at  Instituto Nacional de Astrof\'{i}sica, \'{O}ptica y 
Electr\'{o}nica, Apartado Postal 51y 216, 72000 Puebla,  M\'{e}xico  \email{ vahram@inaoep.mx}}

\maketitle

%\abstract*{
%}

\abstract{ 
%In the presence of non-stationarity, a single, static cross-correlation score for the entire time series is less useful.
Here we investigate  light curves of the  continuum and emission lines  of   five  type 1 active galactic nuclei (AGN) from our monitoring campaign,  to test time-evolution of their
time delays.
%We   show that Gaussian kernel is identical to the cross-correlation (CC) function if covariance of signals is an exponential kernel.
Using both modeled and observed  AGN light curves we  apply  Gaussian-kernel based estimator to capture  variation of local patterns of their time evolving delays. 
%  evolution with time gives insights how relationship among underlying processes is changing.
 The largest variations of time delays of all objects occur in the period when  continuum or emission lines luminosity is the highest.
However, Gaussian kernel based method shows instability in the case of NGC 5548, 3C 390.3, E1821+643 and NGC 4051
 possible due to numerical discrepancies between  Damped Random Walk (DRW) time scale of light curves and sliding time windows of the method. 
The temporal variations of time lags  of  Arp 102B can correspond to the real nature of  the time lag evolution.
%By means of statistical Augmented Dickey Fuller  (ADF) test, we found  strong evidence of   non-stationarity of time series of  NGC 5548 and  3C 390.3.
 % Arp 102B, NGC 4051 and E1821 data sets did  not show  strong conclusive  response to the ADF test.
%We    employed  Gaussian-kernel based estimator to capture  variation of local patterns of their time evolving delays. 
%The extreme variations of time delays of NGC 5548  occurred  during  the global highest and lowest state of activity of its light curves. The abrupt
%change in time lag evolution is recorded in the case of Arp 102B , 3C 390.3, and E1821.In the case of 3C 390.3 it 
 %occurred during the  highest state of activity of  light curves, in the
%case of Arp 102B it corresponds to the prominent feature in emission line, and in the case of E1821 it occurred during the
 %highest state of activity of continuum. NGC 4051, with the shortest monitoring period, where  light curves do not have prominent global highest and lowest activity states,   exhibits diagonal pattern of time delays evolution. 
% These coincidences  imply that  prominent features of time delay evolution of these objects are   triggered  by the  'shocks' in time series  marked as  the highest and %lowest activity states of these objects.
}


\section{Introduction}
\label{sec:1}

%In comparison to some other objects (e.g. X ray binaries), active galactic nuclei (AGN) exhibit much slower variations.
The timescales of  active galactic nuclei  (AGN) optical  variability range from weeks up to years, which causes data analysis problems that
are not usually presented in the case of  the fast variable objects. The large number of observations and good sampling   covering the full range of frequencies
 has almost never been possible to achieve  in AGN observation. AGN time series are obtained from monitoring campaigns  consisting of regular  observations 
over long period, or continuous observations  set on  much shorter time span.
%The basic problem in AGN timing analysis is that we cannot obtain sufficiently long and well-sampled time series.
%Furthermore, these time series are often multivariate, spatially in this work we will consider  two component (bivariate ) time series, consisting of two sets of
% of observations of variations in optical wavebands.
 AGNs  have  strong  emissions lines (e.g. the Balmer line series),
 excited by photoionisation  from the  continuum emission. These lines  respond to   the  variations  of the ionizing continuum, and time delays between continuum and line variations can be related to the  distance (geometry)  between continuum and line emitting regions.
 Since AGN do not vary periodically,  their  delays are seen as ÔechoesÕ of the variations of   the  continuum.
 %The determination of time lags is obscured by the irregular time sampling and  very red  (low frequency dominated)  continuum variations. 
 Also,  the line response to the continuum  can be non-linear, non-stationary at some level.
For  example,  in case  of  NGC 5548 (see Cackett and Horne(2005), and reference therein)  the H$\beta$ lag relative to the continuum  is determined, on a year by year basis 
using  CC, and  it was  found that lag increases with the increase of the  mean continuum flux. Also, various studies shown that emission lines have a nonlinear response to continuum variations (Pogge and Peterson (1992),  Deitrich and Kollatschny (1995)). 

%This 'intrinsic Baldwin effect'
 %(Kinney et al. (1990),  Krolik et al (1991), Koratkar and Gaskell (1991))
%where the H$\beta$ response to variations in the continuum decreases with increasing continuum has been recorded for NGC 5548 
%(Gilbert and Peterson (2003),  Goad et al. (2004)).

Rehfeld and Kurths (2014) investigated similarity estimators  suitable for
 the quantitative measure  of dependencies in irregular and/or non-linear   time series. They introduced    Gaussian-kernel-based CC (captures linear relationship) and Gaussian  mutual information (MI) measure  (captures non linear relationship), and  found that interpolation to regular spacing of the observation times results in worse estimates, while  their adapted estimators are more efficient in the presence of sampling time irregularity. 
  According to their approach it  is possible to calculate  windowed cross similarity between signals and to see how the time delay is changing over the time span covered by observations.

Mostly in CC analysis of AGNs, the underlying hypothesis is that  estimated  quantities  do not vary significantly for the duration
of the  cross-correlated  signals. 
 We investigated  CC and time delays  of continuum and emission line curves of 4 type 1 AGNs: Arp 102B, 3C 390.3, NGC 5548,  and NGC 4051 with assumption that those parameters do not vary with time in our previous work  (see Kova{\v c}evi{\'c} et al. 2014). 
 Our analysis is focused here  on  the  detection of   time evolution  of linear (measured by CC) and non linear relationships (measured by  (MI)) of   time series of  our  data set enlarged with time series of E1821+643.  Some geometrical characteristics of  broad line region (BLR) can be extracted from the time delays , so  any detection of variation of time delays over monitored period can contain information about   its  geometry variations. The structure of the paper is organized as follows.  In the section Data  a short description of used data samples is given. 
In the section Method     Gaussian kernel based CC and  MI method are described. In the section Results and Discussion  the maps of  time delay variations during  observed periods  are displayed 
   and discussed their implications. In the section Conclusion  our main findings are summarized.

\section{Data}
\label{sec:2}
% Always give a unique label
% and use \ref{<label>} for cross-references
% and \cite{<label>} for bibliographic references
% use \sectionmark{}
% to alter or adjust the section heading in the running head
We analyzed the integral fluxes of the broad H${\alpha}$ and H${\beta}$ emission 
lines and the fluxes of the continuum of our objects.
Spectra of Arp 102B, 3C 390.3,  NGC 5548, and E1821+643 are from our monitoring campaign 
Shapovalova et  al. (2004), Shapovalova et al. (2010), Shapovalova et al. (2013), Shapovalova et al. (2015).
As  for NGC 4051 the data were taken from Denney et al. (2006). 
 Detailed data and 
  discussion of observations and data reduction could be found in papers mentioned above.

The sampling pattern (Pessah 2005) is characterized by vector  $SP=(T, T_{S} , T_{B} )$, which components 
 stand for the duration of the observation campaign (T), the sampling interval ($T_S$), and the binning interval ($T_B$).
Since our sets of data are obtained from optical monitoring campaigns, their $T_S=T_{S}(T)$ is  discrete variable over
the time-period of observation. 
We calculated  the rms amplitude  ${\sigma_{rms}}^2$  (Pessah 2005)     associated with each light curve flux L as

\begin{equation}
\sigma _{rms}^2=\frac{1}{P-1}\sum_{n=0}^{P-1}\left | L-\bar{L} \right |^2
\label{eq:01}
\end{equation}
\noindent
  where $P=\frac{T}{\bar{T_{S}}}$,  $\bar{L}$   and ${\bar{T_{S}}}$  are the average values  corresponding to the particular time series and its $T_S$ respectively.
Comparison of   ${\sigma_{rms}}^2$  for all objects as
function of the ratio between the observing and the sampling times is given in Table 1.

\begin{table}
\begin{center}
%\resizebox{5.2cm}{!}{
\begin{tabular}{llll}

\hline

Object & Light curve&$\frac{T}{\bar{T_{S}}}$   & ${\sigma_{rms}}^2$ \\
  
     \hline
& cnt&   115& 2.30\\

{Arp 102B}&      H$\alpha$&   89&  40.38 \\
&             H$\beta$&        117&  2.06 \\

\hline

& cnt&  127 & 1.16\\

3C 390.3&     H$\alpha$&   47&  11.18 \\

&             H$\beta$&        128&  0.87 \\

\hline
&  cnt&   80& 12.37\\

{NGC 5548}&      H$\alpha$&   55&  81.02 \\
&             H$\beta$&     83&  6.11 \\
\hline
NGC 4051 & cnt&   234& 0.19\\

&             H$\beta$&       107&  0.13\\

\hline
E1821+643 & cnt&  126& 1.89\\
&             H$\beta$&     126&  0.52 \\

\hline
\end{tabular}
\end{center}
\caption{  Sampling pattern ($ \frac{T}{\bar{T_{S}}}$) and  the rms amplitude (${\sigma_{rms}}^2$) for continuum (cnt), $H_{\alpha}$ and   $H_{\beta}$ observed light curves.} 

%}
\label{table1}
\end{table}

An example of nonlinearity of our time series is given on the upper panel in the   Fig. 1.  Simple visual inspection of the 6-year long  NGC 5548 light curves
shows that in the lowest state (in 2002) of  the H$\beta$ line  is more concave than the same feature in the  continuum, while in the highest state (1998), the H$\beta$ peak is  not as pronounced as the continuum's peak. Since the  light travel time delay blur  out the emission line response, this blurring  effect  should happen to
both the peak and the concave feature simultaneously. The same pattern  occurred  in 1989-2001 during  AGN Watch monitoring campaign of NGC 5548 (see details in Cackett and Horne (2005)).

%\begin{figure}[h]
%\sidecaption
% For example, with the graphicx style use
%\includegraphics  [scale=0.55 ]{Fig1.eps}
%
% If no graphics program available, insert a blank space i.e. use
%\picplace{5cm}{2cm} % Give the correct figure height and width in cm
%
%\caption{Sampling pattern ($T, T_S$ ) of monitoring campaigns. Time axis  depicts duration (T) of monitoring campaigns, colored vertical lines are  times of %individual observations. Time variable  sampling intervals $T_S(T)$ are void spaces among vertical lines. }
%\end{figure}

\begin{figure}[h]
%\sidecaption
% For example, with the graphicx style use

\includegraphics  [scale=0.4 ]{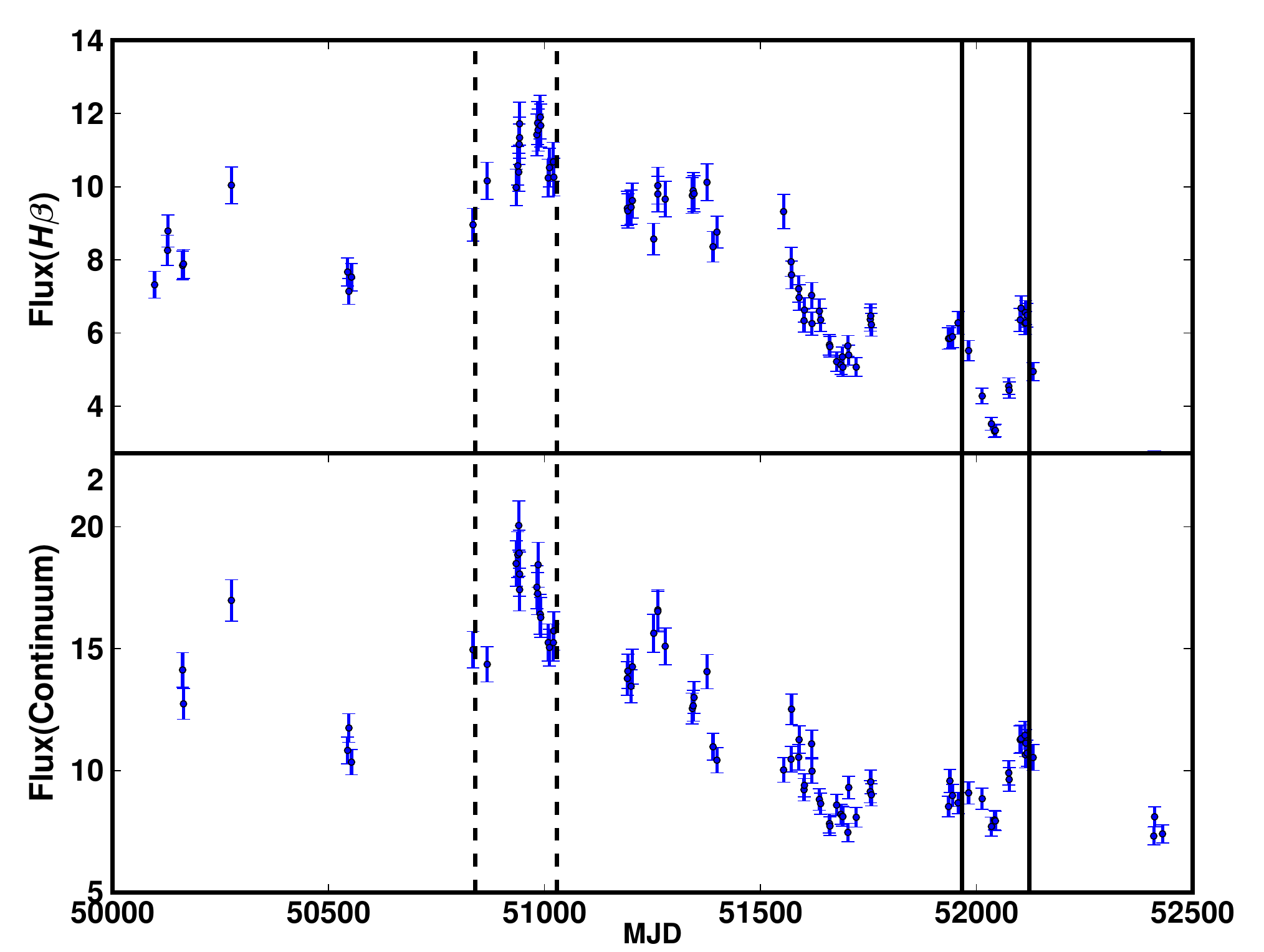}\\ 
\vspace{2.5pt} 
\includegraphics  [scale=0.4]{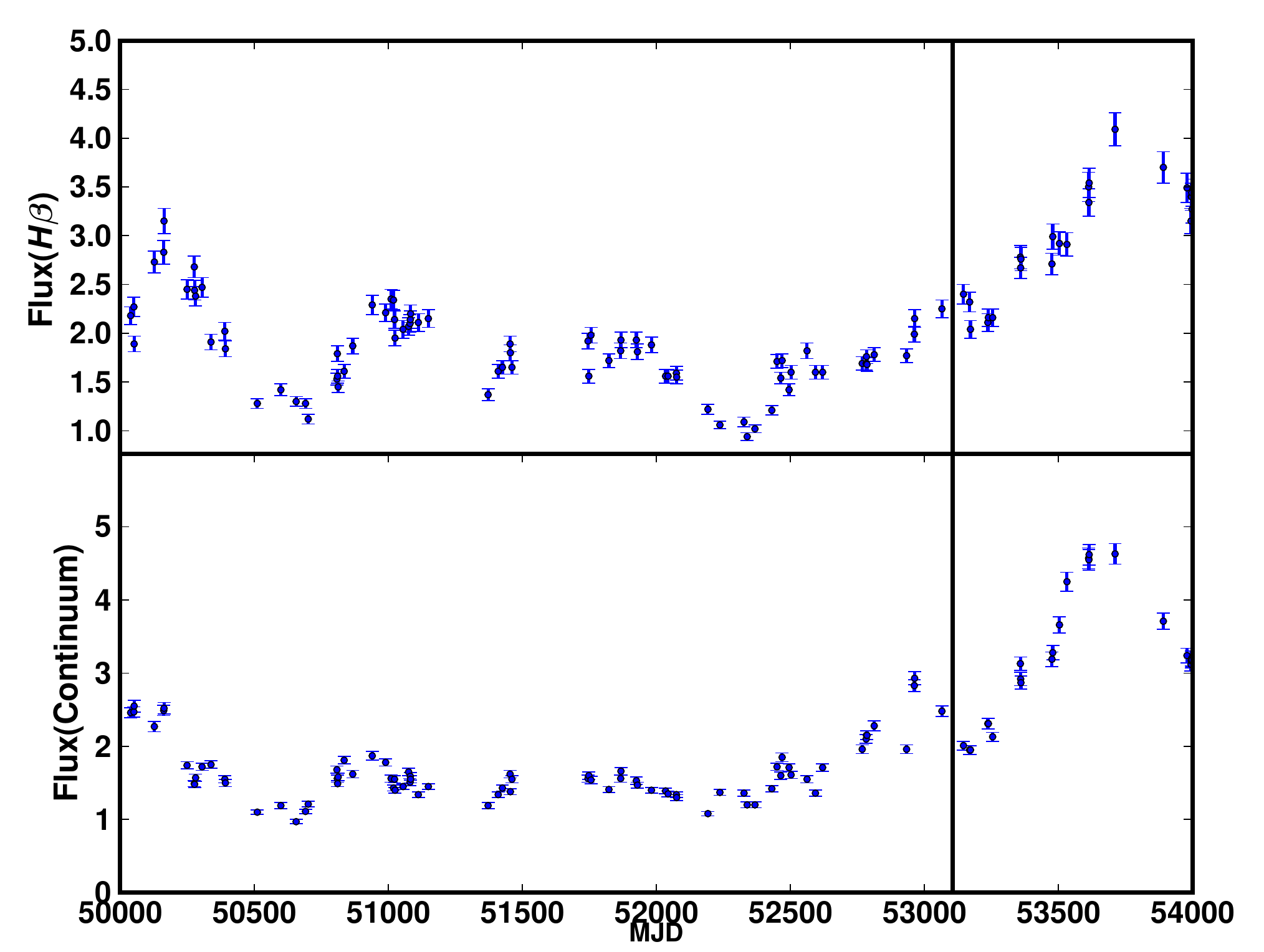}\\ 
\vspace{2.5pt} 
\includegraphics  [scale=0.4]{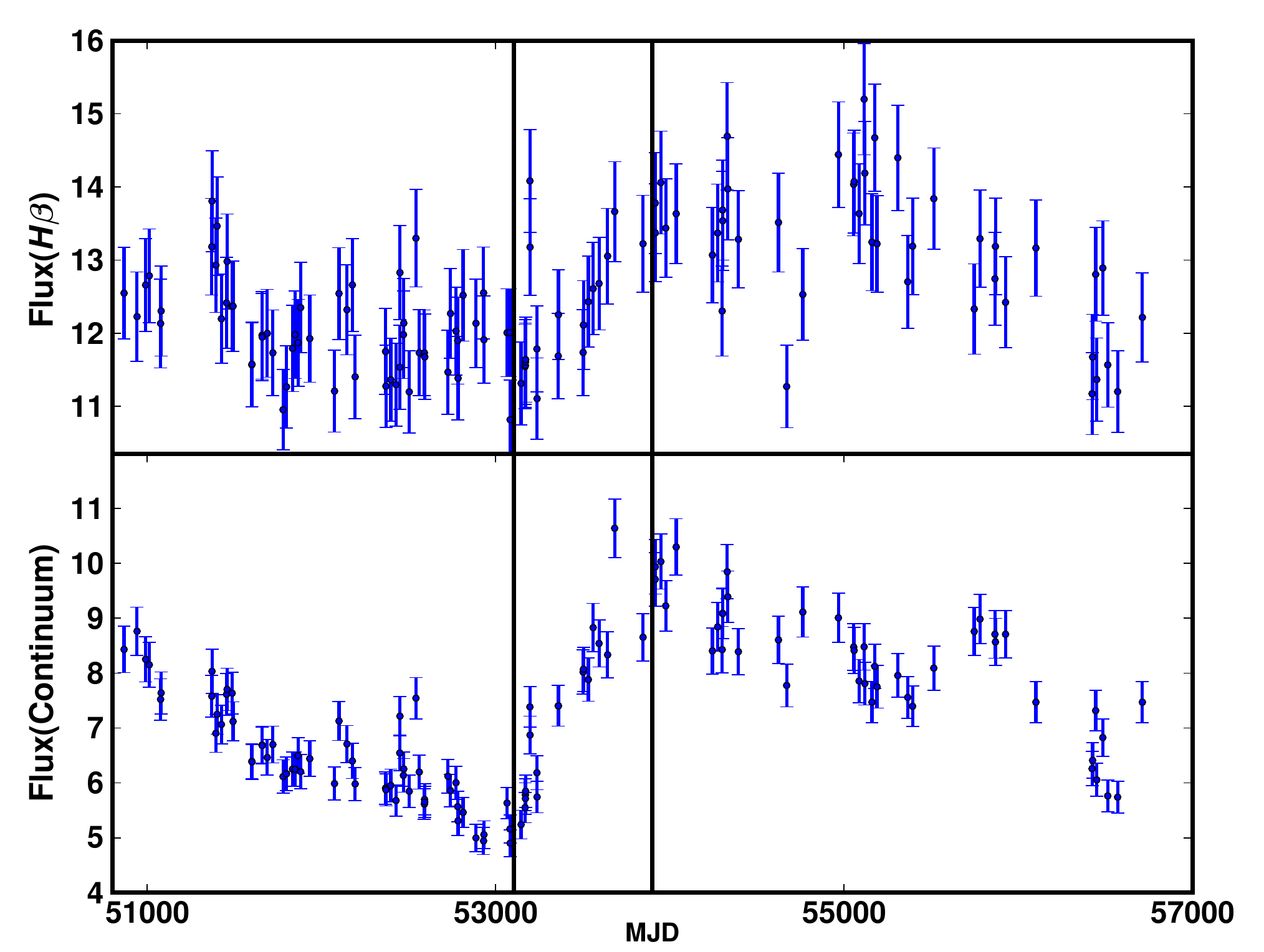}
% If no graphics program available, insert a blank space i.e. use
%\picplace{5cm}{2cm} % Give the correct figure height and width in cm
%
\caption{ Upper plot: Time series  from our spectral monitoring program  data on NGC 5548 for optical continuum flux ( $ 10^{-15} \mathrm{ erg \, s^{-1} cm^{-2}}$  \mbox{\AA} $ ^{-1}$  , at 5190 \AA) and H$\beta$ emission line flux
($ 10^{-13}  \mathrm { erg \, s^{-1}  cm^{-2}} $\AA $ ^{-1}$)  in 1996-2002. Vertical dashed and solid lines mark the highest (1998 Jun 26; JD 2450991) and lowest  (2002 Jun 4; JD 2452430) activity 
 period of the light curves respectively.
Middle  and bottom plot: The same as in upper plot  but for  spectral monitoring  of 3C 390.3 in 1995-2007  and  E1821+643 in 1998-2014. Solid lines mark the highest  activity  period of the light curves.  }
\label{fig:1}       % Give a unique label
\end{figure}

%Use the standard \verb|equation| environment to typeset your equations, e.g.
%
%\begin{equation}
%a \times b = c\;,
%\end{equation}
%
%however, for multiline equations we recommend to use the \verb|eqnarray| environment\footnote{In physics texts please activate the class option \texttt{vecphys} to %depict your vectors in \textbf{\itshape boldface-italic} type - as is customary for a wide range of physical subjects}.
%\begin{eqnarray}
%a \times b = c \nonumber\\
%\vec{a} \cdot \vec{b}=\vec{c}
%\label{eq:01}
%\end{eqnarray}

\subsection{Method}
\label{subsec:2}

From general point of view, the CC (or similarity) allows discovering groups of objects with similar behavior and,
potential anomalies which may be revealed by change in CC.   

Specifically  the time lags between correlated time series of unalike  but related origin reveals  the causalities (linear connections) within 
the system. Namely, the CC coefficients  are calculated based on hypothesis that pairs of values from each series are subjected
to simple linear relationship.
 In order to reconstruct linear dependencies of  hidden processes from the observations, the cross power spectra and CC
 are used due to Fourier transform connection between them (Chatfield (2004)).
 However, the irregularity of  time series sampling  inhibits application of standard methods of  similarity (cross power and CC)
  measures.
    There  are several approaches to overcome this problems,  classified generally  as (Broersen and de Waele (2000)): (i) direct transform methods, (ii) slotting techniques, (iii) model-based estimators, and (iv) time series reconstruction methods.
   The Lomb-Scargle  periodogram (Lomb (1976), Scargle (1981), Scargle (1982)) is a well-known type (i) method that computes a least squares fit of sine curves to the data.  Lomb-Scargle is powerful in finding narrow peaks with a low noise level.  However, it loses its strength with smaller peaks and more noise.

   Standard    slotted CC estimation discretizes the distance between two observations to slots of 
width  $\Delta$. 
The product of two irregular observations contributes to the slotted autocorrelation at a certain lag $ k \Delta$,
if their distance is within the range $(k \pm 0.5 )\Delta $  (Mayo (1993),  Edelson and Krolik (1988)).
In such a way,  this technique interpolate in time delay domain (contrary to the  interpolation in the time domain 
for one or both members of the correlation (see Gaskell and Sparke (1986), Korista and Goad (1991)).
The disadvantage of this technique is  the correlation function estimates are not necessarily positive semidefinite and the spectra computed from their Fourier transform can show negative power.  So some kind of post-processing techniques are introduced to avoid this problem (Stoica et al. (2008), Babu et al. (2009)).

Model-based estimators fit a model to the time series and then try through some kind of transformation to determine the time lag between time series, such example
is SPEAR method (Zu et al. (2011)).
The fourth group of estimators consists of  different kind of interpolation methods in time domain of time series.
    The case of  the standard linear interpolation of  irregular observations, onto an 
  even sampling grid,  adds bias towards low frequencies in power spectral density (PSD) estimation. 
  Application of cubic spline interpolation can lead to spurious peaks, so the variance of resampled signal 
  is too high.  As opposed to cubic spline,  linear interpolation is a robust resampling method, it does not
  add spurious peaks, but the variance of resampled signal is too low. This arises from the fact that 
  linear interpolation is weighted average of  two irregular observations, so the resampled signal will have lower
  variance than original one.
  
Rehfeld  et al. (2011) probed  different CC estimators for irregular time series and found that a
   Gaussian-kernel-based estimator performed best,  which has the    form  introduced in  Bjoernstad and Falck (2001) :

\begin{equation}
ker(k-\Delta t_{ij}^{xy})=ker(d)=\frac{1}{\sqrt{2\pi h}}exp^{\frac{-\left | d \right |^2}{2h^2}}
\label{eq:02}
\end{equation}

\noindent 
where $\Delta t_{ij}^{xy}$ is the  inter-observation time, and k denotes the k-th lag.
 The CC function can be written as

\begin{equation}
CC(k \Delta \tau)=\frac{\sum_{i=1}^{N_{x}}\sum_{j=1}^{N_{y}} x_{i} y{j} ker}{\sum_{i=1}^{N}\sum_{j=1}^{N}ker}
\label{eq:03}
\end{equation}

\noindent 
where $x_{i}, y_{j}$  and  $N_{x}, N_{y}$    are time series values and their lengths respectively, $k \Delta \tau$ is k-th time lag.
Gaussian window has three  desirable properties:(i) optimal resolution.  Only the Gaussian window minimizes the resolution product of its width and width 
of its Fourier transform. (ii) Isotropic (circularly symmetric). Isotropy implies no directional bias in multi-dimensional correlations. (iii) Separable in N dimensions.
It implies computational efficiency. (iv) the product of two Gaussians with equal widths is Gaussian.

In our analysis we are interested to track any possible time-evolving presence in our data, by means of time-evolving correlation function.
Rehfeld method allows  splitting signals into overlapping segments and calculating time-evolving Gaussian kernel CC.

On the other hand we know that the covariance functions of the AGN light curves are close to
exponential kernels, i.e, the damped random walks (DRW), on scales beyond a week, and probably have
a break on smaller scales based on Mushotzky et al. (2011) analysis of the Kepler mission light curves.
 It has been  confirmed by Kelly et al. (2009), Kozlowski et al. (2010), Macleod et al. (2010),  Zu et al. (2013), Andre et al. (2013).

Also,  we are interested to measure any possible non-linear correlations among our time series (the case of NGC 5548 is known as non-linear, see  Fig. 1). 
 $MI$ can  measure both linear and nonlinear relationship between time series.

  MI strictly  determines  (in units of bits)  how much information the value of one variable reveals about the value of another.
 This is of great  importance  in information theory (Cover and Thomas (1991)). 
 
 Perhaps less well known,  is applicability of  MI on the statistical analysis of data.  It is not strange since  the mathematical theory of information transmission established by Shannon (1948) represents the culmination of statistical research.  Namely the ideas from statistics  become   applicable to a strikingly  wide range
  of  problems.
To get an intuitive sense of MI notion,  let us consider an  example of  two {\it independent} random variables X, Y.  Since they are independent,
their joint distribution $p(X,Y)$ is identical to the product of  marginal distributions $p(X)$ and $p(Y)$.
So one could investigate the degree of independence between X and Y by computing probabilistic distance 
between $p(X)p(Y)$ and $p(X,Y)$. This distance in the case of two independent variables is 0.
A common probabilistic distance between variables are measured by Kullback-Leibler divergence, and if we apply it 
to our intuitive example to measure distance between the joint distribution and the marginals of two random variables
we will come to the concept of MI inevitably.
  The MI  (I(X,Y)) between the   variables $X=(x_{1},...x_{n})$ and 
$Y=(y_{1},...,y_{m})$ is defined (Shannon (1948), Kolmogorov (1968)) in terms of entropy 
($H(X)=-\sum_{i=1}^{n}p(x_{i})\log p(x_{i})$) as

\begin{equation}
I(X,Y)= H(X)+H(Y)-H(X,Y) \geq 0
\end{equation}
 
 or in terms of their joint probability distribution p(X,Y) as
 
  \begin{equation}
I(X,Y)=\int dx dy p(x,y) \log_{2} \frac{p(x,y)}{p(x)p(y)}
\end{equation}
MI definition assures that its value is always nonnegative and zero
only when $p(X,Y)=p(X)p(Y)$. The MI will be greater than zero when X and Y are dependent mutually, regardless
of  the level of  nonlinearity  of that  relationship. The stronger mutual dependance implies the larger the value 
of I(X,Y).
One can say that MI is deeply connected to the statistical problem of detecting dependencies. Namely, from previous equation,
it is clearly seen that for the data taken from distribution p(X,Y),  I(X,Y) quantifies
the expected log-liklihood ratio of the data with underlying p(X,Y) as opposed to $p(X)p(Y)$.
So $I(X,Y)^{-1}$  is the typical amount of data needed to be  collected  in order to 
make double increase in posterior probability of variables dependance  relative to the hypothesis 
that they are independent.  On the  other hand Neyman-Pearson lemma (Neyman-Pearson (1933)) asserts that  all useful information
about differentiating between two hypothesis is contained in 
$\sum_{i}\log_{2}\frac{p(x_{i},y_{i})}{p(x_{i})p(y_{i})}$,   possessing the maximal possible statistical power
for such test. Based on this conclusion inferred from  Neyman-Pearson lemma, one can say that 
 $I(X,Y)$ (defined to be exactly the limit of such sum)   provides upper constrain  on how well can be performance of any dependance test applied on data obtained
 from distribution $p(X,Y)$. 
Estimation of MI is nontrivial, because of nontrivial estimation of the joint distribution
p(X,Y) from a finite sample of N observation. There are several approaches to overcome this problem, and Rehfeld and Kurts (2014) implemented one 
of them.
  Similar to the Gaussian kernel based  CC, Rehfeld and Kurths (2014) introduced  Gaussian kernel based MI. 
 Algorithmically, the procedure consists of constructing a new  bivariate set of observations
 ($Y^{X}$) by estimating  for each point in time series $X=({t_{i}^{x}}, x_{i})$ a local Gaussian-kernel-based weighted mean in second time series 
 $({t_{i}^{y}},y_{i})$.
 The same procedure is repeated for the first time series, obtaining $X^{Y}$. Then original and reconstructed series are concatenated
  $(X\cup X^{Y}),(Y\cup Y^{X})$. The joint density of X and Y is estimated using standard binning estimators for  MI. This procedure
  also can track time evolving of  MI.

\subsubsection{Results and discussion}

There is a large theoretical effort  to understand the innermost part of active nucleus parametrized by  black hole mass, spin and Eddington ratio.
 However, this process is not stationary imposing additional problems (Czerny and Hryniewizc (2012)).
 All our time series (see Kova{\v c}evi{\'c} et al. (2014) and reference therein) are very fluctuated, with structural change. 
Therefore,  we   firstly  examine  properties of  our  data sets by using   stationarity test.
 Time series stationarity is a statistical characteristic of a seriesÕ mean and variance over time. 
If both are constant over time, then the series is assumed to be a stationary process 
(i.e. is not a random walk/has no unit root), otherwise, the series is described as being a non-stationary process 
(i.e. a random walk/has unit root).

Stationarity of a series is an important phenomenon because it can influence its behavior.
  For example, the term ÔshockÕ (Brooks (2014)) is used frequently to indicate an unexpected change in the value of a variable (or error).
   For a stationary series a shock will gradually terminate. That is, the effect of a shock during time t 
 will have a smaller effect in time $t+1$, a still smaller effect in time $t+2$, etc.

Non-stationarity  can raise from deterministic changes  like trend or seasonal fluctuations or the stochastic properties of processes.
  In the case of  testing for non-stationarity and stationarity of the time series 
autoregressive unit root tests  are valid if the time series $y_{t}$
 is well described by an AR(1) with white noise errors.  However, many time series
 posses a complex dynamic structure which cannot  be captured by a simple AR(1) model. 
  Said and Dickey (1984) augmented the basic autoregressive unit root test to allow for general 
 ARMA(p,q) models with unknown orders which is called the Augmented Dickey- Fuller (ADF) test. 
 The ADF test probes  the null hypothesis that a time series is a non-stationary against the alternative that it is stationary 
 assuming that the underlying data dynamics  has  an ARMA structure. 
 
 The ADF test  estimates the test regression in the form
 
 \begin{equation}
 y_{t}={\bf D_{T}} +\phi y_{t-1}+\sum_{j=1}^{p} \psi_{j} \Delta y_{t-j} +\varepsilon_{t}
 \end{equation}
\noindent where ${\bf D_{T}}$ is a vector of deterministic terms (constant, trend, etc).  Imposing constrains
on constant and time trend to be zero corresponds to modeling a damped random walk, while
setting only time trend to be zero corresponds to modeling a random walk with a drift. The total of p lagged difference terms, $\Delta y_{t-1}$
approximate the ARMA structure of the errors.  The value p is set  that the error $\varepsilon_{t}$ is serially uncorrelated.
The error term is assumed to be homoskedastic.  The unit root test is then carried out under the null hypothesis $\phi=0$
against the alternative hypothesis of $\phi<0$. The ADF t-statistic ($ADF_{t}$) 
 and  normalized bias statistic ($ADF_{N}$) are calculated by means of  the least squares estimates of
equation and are given by

\begin{equation}
\begin{array}{rcl}
ADF_{t}=t_{\phi=1}=\frac{\check{\phi}-1}{SE(\phi)} \\
ADF_{N}=\frac{N(\check{\phi}-1)}{1-\check{\psi_{1}}-\cdot\cdot\cdot -\check{\psi_{p}}}\\
\end{array}
\end{equation}

\noindent where $\check{\phi}$ is an estimate  of the coefficient $\phi$, $SE(\phi)$ is its standard error in alternative model, 
$\check{\psi_{1}}\cdot\cdot\cdot \check{\psi_{p}}$ are stationary coefficient in alternative model, and N is the effective sample size.

We employed  Augmented Dickey-Fuller (ADF) test implemented in SciKits Statmodels of time series analysis in Python language.
 We tested our time series   with a null hypothesis of non-stationarity  against an alternative hypothesis  of stationarity around:constant, a constant $+$ trend, a constant$+$ trend $+$ trend squared,  and no constant trend respectively.
 We summarize  in Table 2 outputs of testing our time series around no constant trend,  which is equal to modeling damped random walk, as we mentioned above.
 When  we compare the test statistics and critical value, we have to reject the null
 if test statistics  is smaller than critical value (not absolute value,  because that is usually applied to two-tailed test).
For example if  the test statistics  is  $-1.85$  and critical value is $ -3.5$  we will not reject the null hypothesis.

Time series of NGC 5548, 3C390.3, H$\beta$ line of NGC 4051 and continuum (5100 \AA)   of E1821+643 are clearly non stationary in all tests at  all confidence levels. 
While for  other time series the non-stationarity is clearly  expressed when testing them without constant trend.
We accepted all our light curves as non-stationary, since their ADF test values  around no constant trend confirm their non-stationarity.
Due to non stationarity CC should  evolve with time, which we  investigated by techniques of Gaussian based CC and MI.  The application of these techniques  on our data sets is facilitated by  a Matlab toolbox (Rehfeld and Kurths (2014),  http://tocsy.pik-potsdam.de/nest.php), modified for our purposes.
The CC coefficients are estimated for each lag shift given  in the
input vector. The signals are splitted into overlapping segments, both splitting window (W) and overlapping parameter (O) must be specified.
Within the overlap of the individual pairs,
 we calculate  the matrix  with their CC coefficients for each lag as specified by input vector of lags.
 The matrix contains N rows (N is the length of input lag's vector) and $k=numpy.floor (ceil)((T-W)/((1-O)*W)+1)$ columns, where
 T is the length of the signals (in time units).

We applied Gaussian kernel with width parameter  0.25,  overlap 0.0-0.3 and used sliding windows between 200 and 500 days.
Fig. 2  depicts how the CC  time lags develop over monitoring time.
NGC 5548 both lines  shows continuous    symmetrical change of CC time lags over time, with larger amplitude in the case of the H$\beta$ line.
The largest and smallest variation for both lines occurred within time window corresponding to the highest state and lowest state of activity (compare Fig.1 and Fig.2).
However,  we find that  MI method applied  on  both lines of  NGC 5548 (Fig.7),  as well as for other cases,  fails, as it only infers the results for time windows with enough points 
(dark color in the Fig.7).

Contrary  to the NGC 5548, the both lines of Arp 102B and H$\beta$ lines of 3C 390.3  and E1821+643 exhibit abrupt 'regime shift':  from nonsignificant lags ( several days) up to about magnitude of hundred days. The magnitude of shift is larger in the case of H$\beta$ than H$\alpha$ line of  Arp 102B, while in the case of
3C 390.3   and E1821+643  we did not have enough data for  H$\alpha$ line and  H$\gamma$  analysis, since it has been required by method at least of 50 points in each sliding window.

 It is clearly seen  that abrupt change in time evolution of  3C 390.3 lags (Fig.2) corresponds to the highest state of  both  light curves (Fig. 1), while in the case of
  E1821+643, the regime shift corresponds to the highest state of continuum only. In the case of Arp 102 B 'the regime shift' of time lag evolution corresponds to the prominent feature of the H$\beta$  line around  MJD 54000  (see Fig.3 in  Shapovalova et al. (2013)).

In the case of NGC 4051, the almost diagonal patches  are occupied  displaying tendency that time lag evolve linearly over time.  In the case of other time series, the absence of such pattern could display tendency  for non-linear time delay evolution.
Both light curves of NGC 4051 (see  Fig. 3 in  Denney et al. (2009)) display almost oscillatory behavior without prominent changes.  

  To obtain a quantitative understanding of the feasibility  of above mentioned results,  we modeled a five artificial  cases using synthetic
light curves obtained by the Cholesky decomposition technique implemented
in JAVELIN code (Zu et al.  2011).
 One virtue of JAVELIN is that it produces an explicit
mean model light curves  constrained by the data.

 Since  the structures shown in the Fig.  2 (c, d panels) exhibits strange {\bf behavior}, the lags change from something close to zero to a
hundred days,  we constructed  10 to 30 times denser  synthetic light curves than original ones by  taking into account the length of monitoring baselines corresponding to original length of monitoring campaigns  and cadences of  daily  or lower value.

For each case we generated 50 MCMC realizations of light curves assuming  that fictive objects 
posses  variability parameters of our 5 objects: variance  $\sigma [mag]$ and DRW time 
   scales of $\tau_{D}[days]$. The panels in the Fig.  3   and Fig. 4 show the Case 1, 2, 3 and Case 4, 5  mean model light curves respectively  while 
   the  results of application of  Gaussian kernel based method
 on those synthetic light curves can be seen in the Figs.  5 and 6. Time evolved Gaussian kernel  lags (c, d panels in  the Fig. 5)
    of synthetic light curves  (see the middle panel in the  Fig. 3)   cloned from the mother curves of Arp 102B, 
    do not exhibit such changes as those seen in the  Fig. 2 (c and d panels). {\bf  Clearly, in the case of Arp 102B clones, temporal non variability of time lags  is a consequence 
    of  stationarity and  constant lags of those curves.
    However, other cases of cloned light curves show strong variation in lags.
   In the case cases of cloned light curves of NGC 5548, 3C390.3,
E1821+643  we notice that  DRW time scales are larger  than used sliding time windows of the Gaussian method, while in the case of NGC 4051 those values are close to each other  ($\tau_D=3.8$ and $W\sim 10$).
These anomalies can rise the instability of Gaussian kernel based method which induces significant variations of time lags of those cloned curves.
 From the other hand it seems that DRW time scales of Arp 102B light curves, have 'Goldilocks' values, and  the method  shows more stability.}

Moreover, all synthetic cases exhibit  the extremum  values of time lags within the time range corresponding to   the shocks in time series.

The static time lags derived by standard slotting  and model technique (ZDCF and  SPEAR, Kova{\v c}evi{\'c} et al. (2014))  are given in Table 3. The magnitude of  lag time evolution for all our cases over monitoring periods are within $3 \sigma$ interval of values obtained by  standard techniques (Table 3).

 Since many  authors reported individual time lag  determinations of our objects (compiled   in  Kova{\v c}evi{\'c} et al. (2014))
 with the exception of E1823+643, we were able to calculate  their weighted mean  in order to have a more informative comparison to the evolved
time lags. As the time lags are accompanied   with asymmetric errors, their weighted mean and error combinations
can  be found  by Model 1 reported in  Barlow (2003). This Model provides convergence of iterative  numerical solution. 
If values and their asymmetric errors are given as $(x_{i},{\sigma_{i}^{-}}, {\sigma_{i}^{+}})\in f_{i}$, where $f_{i}$
is a distribution, the combination of errors is done by finding the variance 
$ V_{i}= 0.25(\sigma_{i}^{+}+\sigma_{i}^{-})^{2}+0.25(\sigma_{i}^{+}-\sigma_{i}^{-})^{2}(1-\frac{2}{\pi})$ 
 and skew 
 $ \gamma_{i} =\frac{1}{\sqrt{2\pi}}[2({\sigma_{i}^{+}}^{3}-{\sigma_{i}^{-}}^{3}) -\frac{3}{2}(\sigma_{i}^{+}-\sigma_{i}^{-})({\sigma_{i}^{+}}^{2}-{\sigma_{i}^{-}}^{2})+\frac{1}{\pi}(\sigma_{i}^{+}-\sigma_{i}^{-})^{3}] $
  for each contributing distribution separately. Adding these up gives the cumulants V (variance) and $\gamma$ (skew) of the combined
  distribution function. If we write $D=\sigma^{+}-\sigma^{-}$ and $S={\sigma^{+}}^{2}+{ \sigma^{+}}^{2}$ then
   the combined errors $\sigma^{+}, \sigma^{-}$ are obtained 
  numerically from following nonlinear equations
  
  \begin{equation}
  \begin{array}{rcl}
  S &=& 2V+\frac{D^2}{\pi}\\
  D & =& \frac{2}{3S}(\sqrt{2\pi}\gamma- D^{3}(\frac{1}{\pi}-1) \\
  \end{array}
  \end{equation}
  
  From the equations above one can find the solution for D and S, after that simple algebraic manipulation leads to the  $\sigma^{+}, \sigma^{-}$.
As for an unbiassed weighted mean estimator one has to take following expression

\begin{equation}
\check{x}=\frac{\sum_{i=1}^{n}\omega_{i}(x_{i}-b{i})}{\sum_{i=1}^{n}\omega_{i}}
\end{equation}

\noindent  where $b_{i}=\frac{\sigma_{i}^{+}-\sigma_{i}^{-}}{\sqrt(2\pi)}$ and weights
are given as  $\omega_{i}=\frac{1}{V_{i}}$.
 We calculated  the weighted mean of time lags  obtained by other authors ($\check{\tau}$) and combined their errors in the form
 of $\sigma^{+}$ and $\sigma^{-}$ according
 to Barlow method  (see Table 4).
 Their values
 agree with  those time evolved. Namely, the time evolved lags 
 $\tau $ are within the region $[\sigma^{+}+\check{\tau}, \sigma^{-}-\check{\tau}]$ of weighted mean lags.
   There is only slight difference in the case of time lags of  Arp 102B.

\begin{table}

\begin{center}

%\resizebox{5.2cm}{!}{
\begin{tabular}{llllll }

\hline

Object &LC&      $  TS_{nc}$  &    nc ($1 \%$)  & nc($5 \%$)& nc($10 \%$) \\

\hline

\multirow{2}{*}{E1821+643}  & { cnt}   &   -0.17 &   -2.6& -1.94 &-1.6\\
                                    
      &{  H$\beta$   } &        -0.15 &   -1.94 &-1.94& -1.61\\
   \hline                                      
\multirow{3}{*}{Arp 102B}&  {cnt}&    -0.42     &   -2.6  &-1.9&-1.6\\
           
   &{ H$\alpha$}   &        -0.95 &   -2.6&-1.9&-1.6\\

 &{ H$\beta$}  &      -0.72 &  -2.6&-1.9&-1.6\\

\hline
\multirow{2}{*}{NGC4051} &   {cnt}  &     -0.6&  -2.6&-1.9&-1.6\\
                                           
                           & { H$\beta$} &    -0.4& -2.6&-1.9&-1.6\\
 \hline
\multirow{2}{*}{3C 390.3}&  {cnt}     &       0.8&   -2.6&-1.9&-1.6\\

   & { H$\beta$}        &       0.3 &  -2.6&-1.9&-1.6\\
\hline
\multirow{3}{*}{NGC 5548}  & {cnt} &       -0.650   & -2.6&-1.9&-1.6\\

  & { H$\alpha$}  &   -1.3  &      -2.6&-1.9&-1.6\\

  & { H$\beta$}    &    -1.5 &  -2.6&-1.9&-1.6\\

  \hline
\end{tabular}
\end{center}
%\hspace{0.5in}\parbox{10in}{\caption{The Augmented Dickey-Fuller statistics for the test of   non stationarity.  LC-Lightcurves; $ TS_{c}$, $ TS_{ct}$, $ TS_{ctt}$ , $ %TS_{ctt}$ are test statistics
%involving constant term, a constant $+$ trend, a constant$+$ trend $+$ trend squared, no constant trend respectively; c,ct,,ctt,nc are critical values for the test %statistics at the $1\%$, $5 \%$, $10 \% $  levels; NS - logical values of non stationarity test.The  null  hypothesis that process is non-stationary is rejected
 %if test statistics  is smaller than critical value (not in the sense of their  absolute values, because that is usually applied to two-tailed test). } }

%}
\label{table2}
\end{table}

%\addtocounter{table}{-1}

\begin{table} [t!]
\caption{The Augmented Dickey-Fuller statistics for the test of   non stationarity. In the column  LC are given light curves;  $ TS_{nc}$ column provides test statistics
involving  no constant trend ; the column nc gives critical values for the test statistic at the $1\%$, $5 \%$, $10 \% $  levels. In all cases the null-hypothesis 
that process is non-stationary is satisfied.}
% NS - logical values of non stationarity test.The  null  hypothesis that process is non-stationary is rejected
% if test statistics  is smaller than critical value (not in the sense of their  absolute values, because that is usually applied to two-tailed test). } 
\end{table}

\begin{figure}[htb]
\captionsetup[subfloat]{farskip=1pt,captionskip=1pt}
\centering
  \subfloat[NGC 5548  H$\alpha$]{%
    \includegraphics[width=0.49\textwidth]{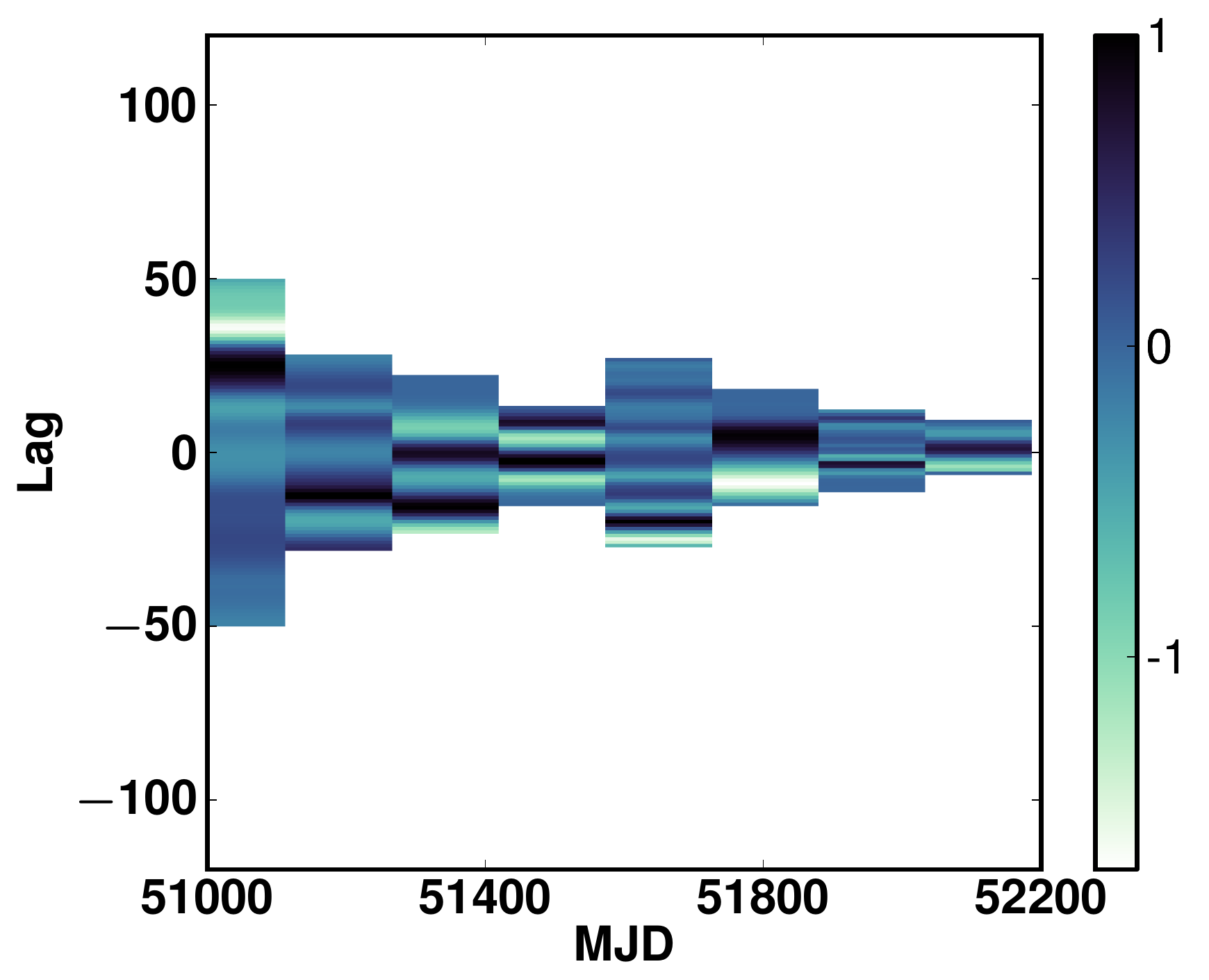}}\hfill
  \subfloat[NGC 5548  H$\beta$]{%
    \includegraphics[width=0.49\textwidth]{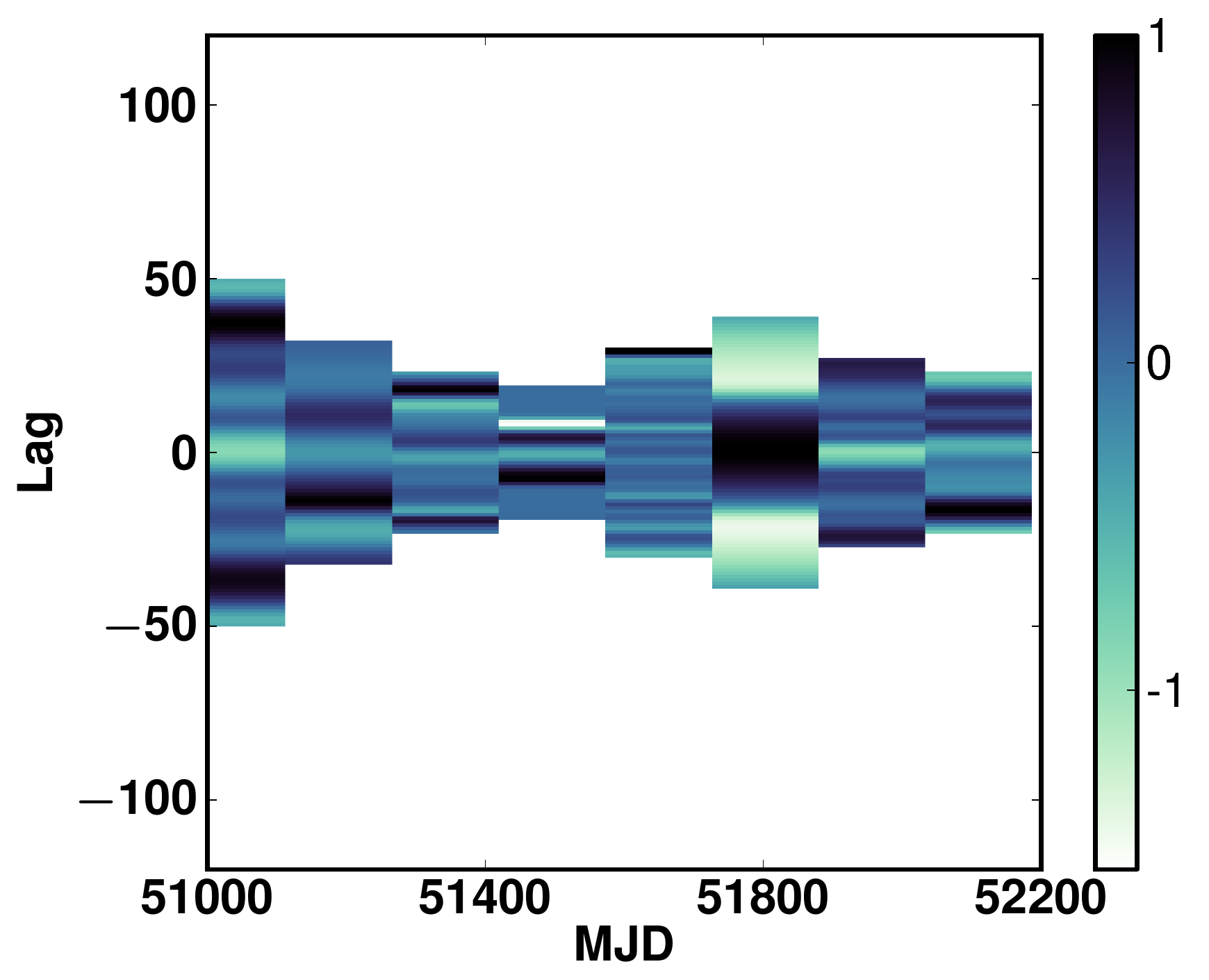}}\hfill\\
    \subfloat[Arp 102B  H$\alpha$  ]{%
    \includegraphics[width=0.49\textwidth]{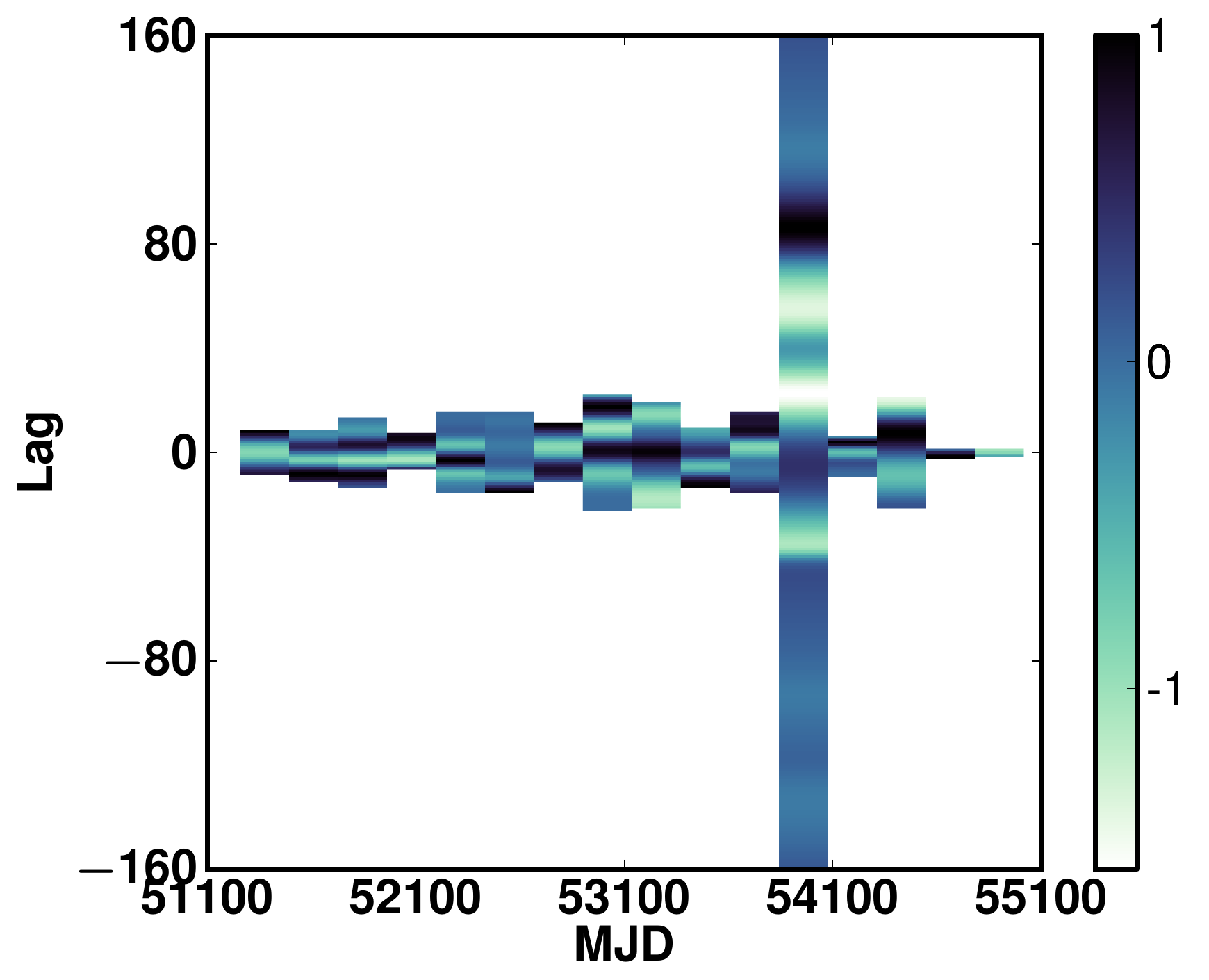}}\hfill
  \subfloat[[Arp 102B  H$\beta$  ]{%
    \includegraphics[width=0.49\textwidth]{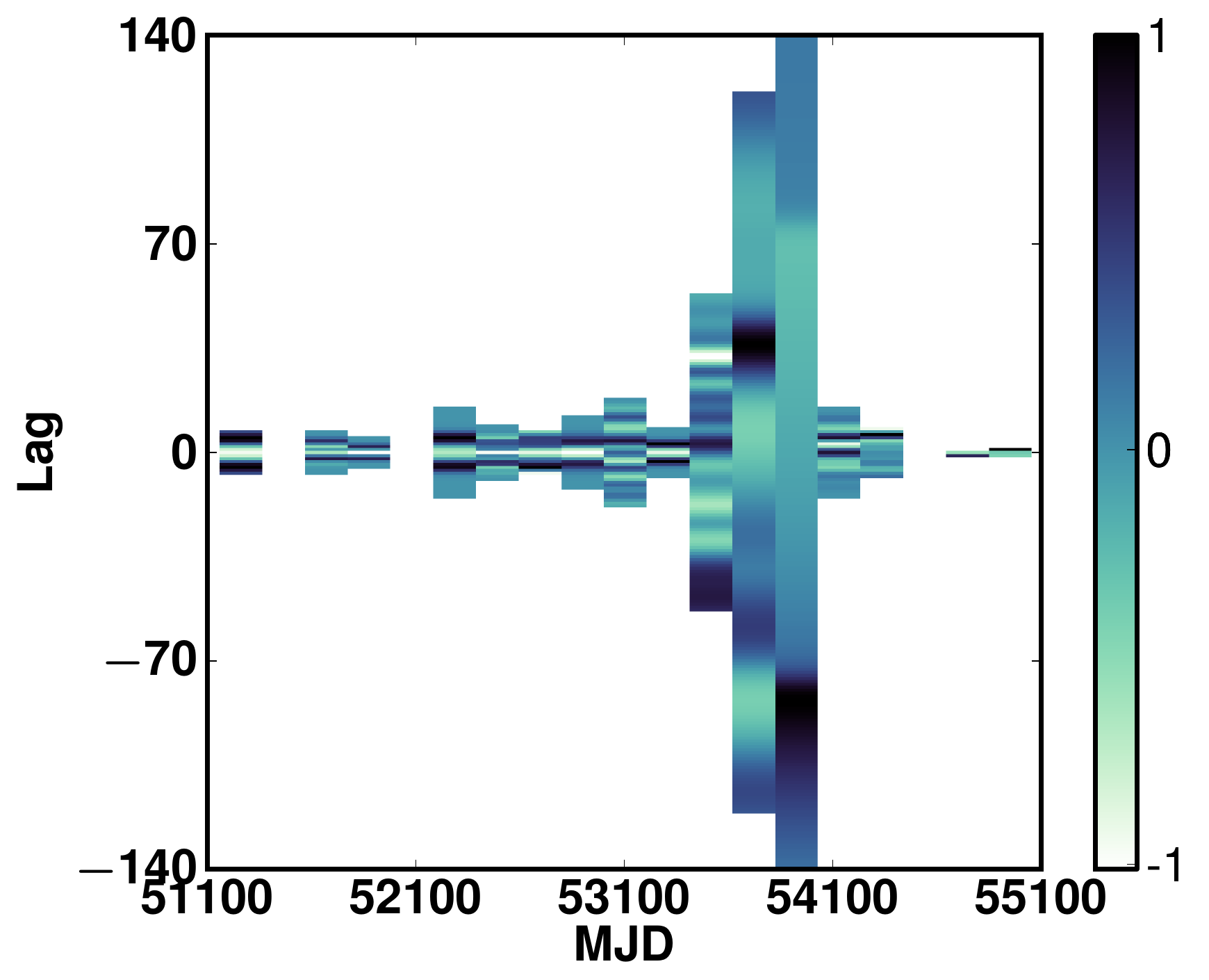}}\hfill \\
      \subfloat[NGC 4051  H$\beta$   ]{%
    \includegraphics[width=0.49\textwidth]{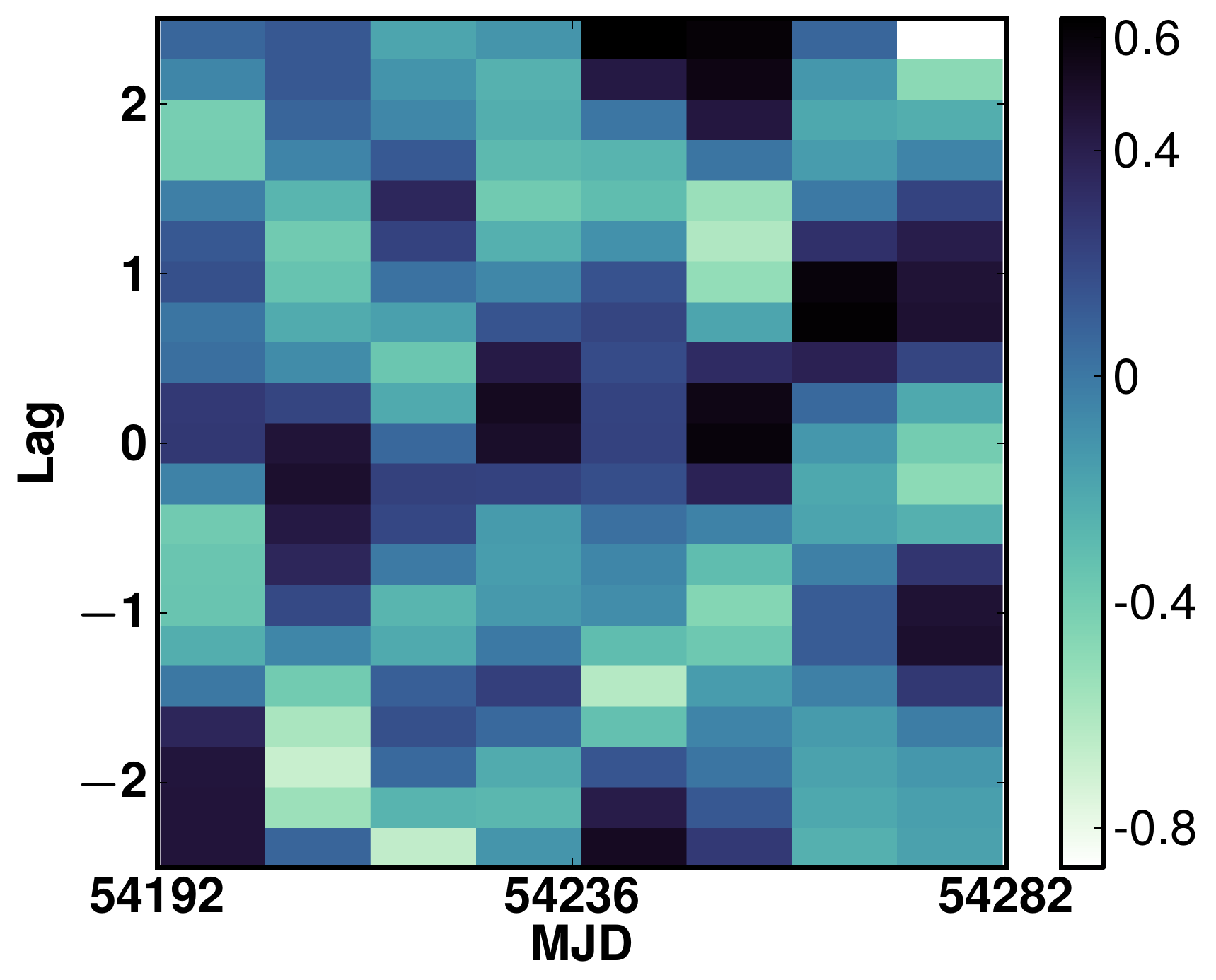}}\hfill
  \subfloat[3C 390.3  H$\beta$   ]{%
    \includegraphics[width=0.49\textwidth]{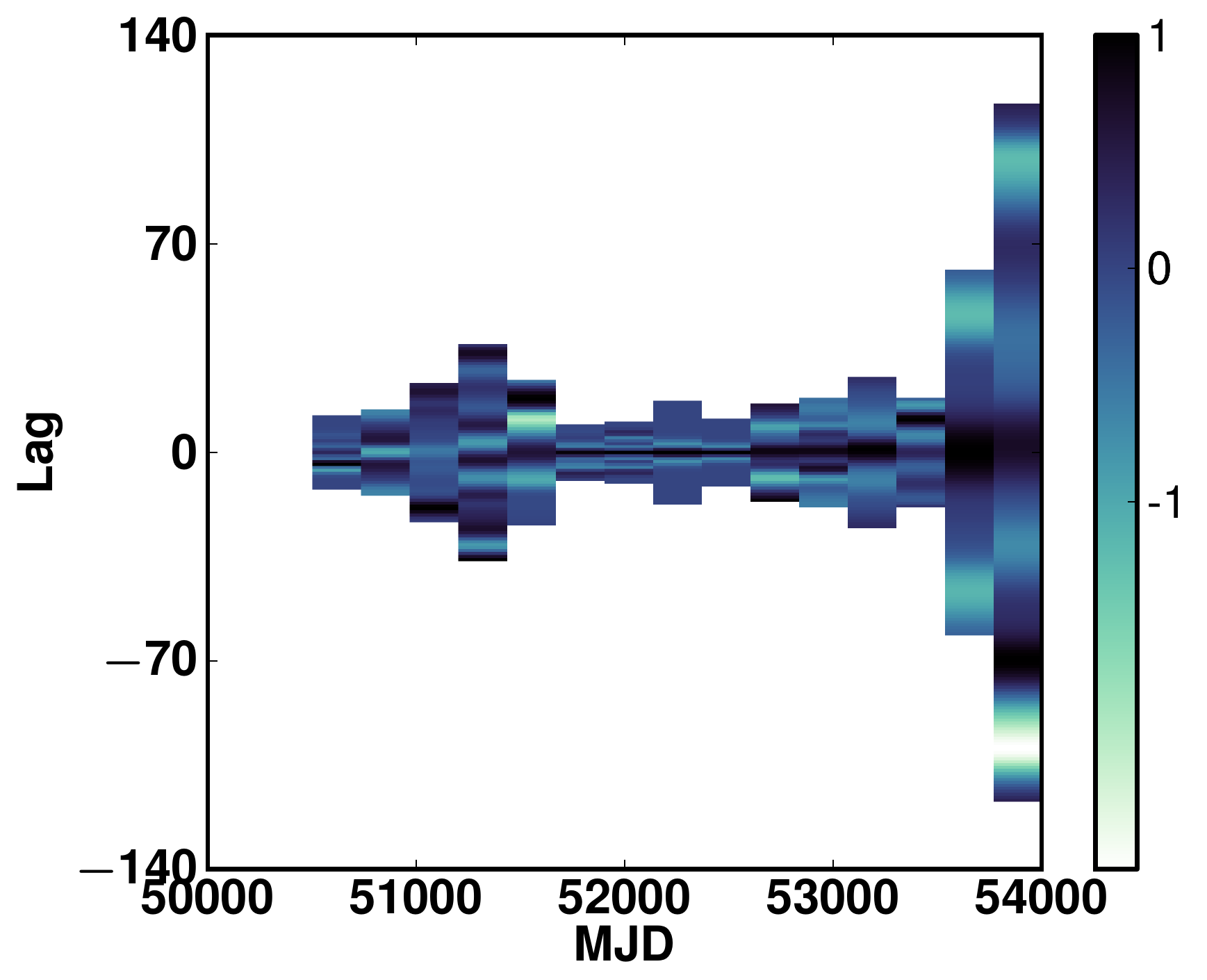}}\hfill
  \subfloat[E1821+643 H$\beta$ ]{%
    \includegraphics[width=0.49\textwidth]{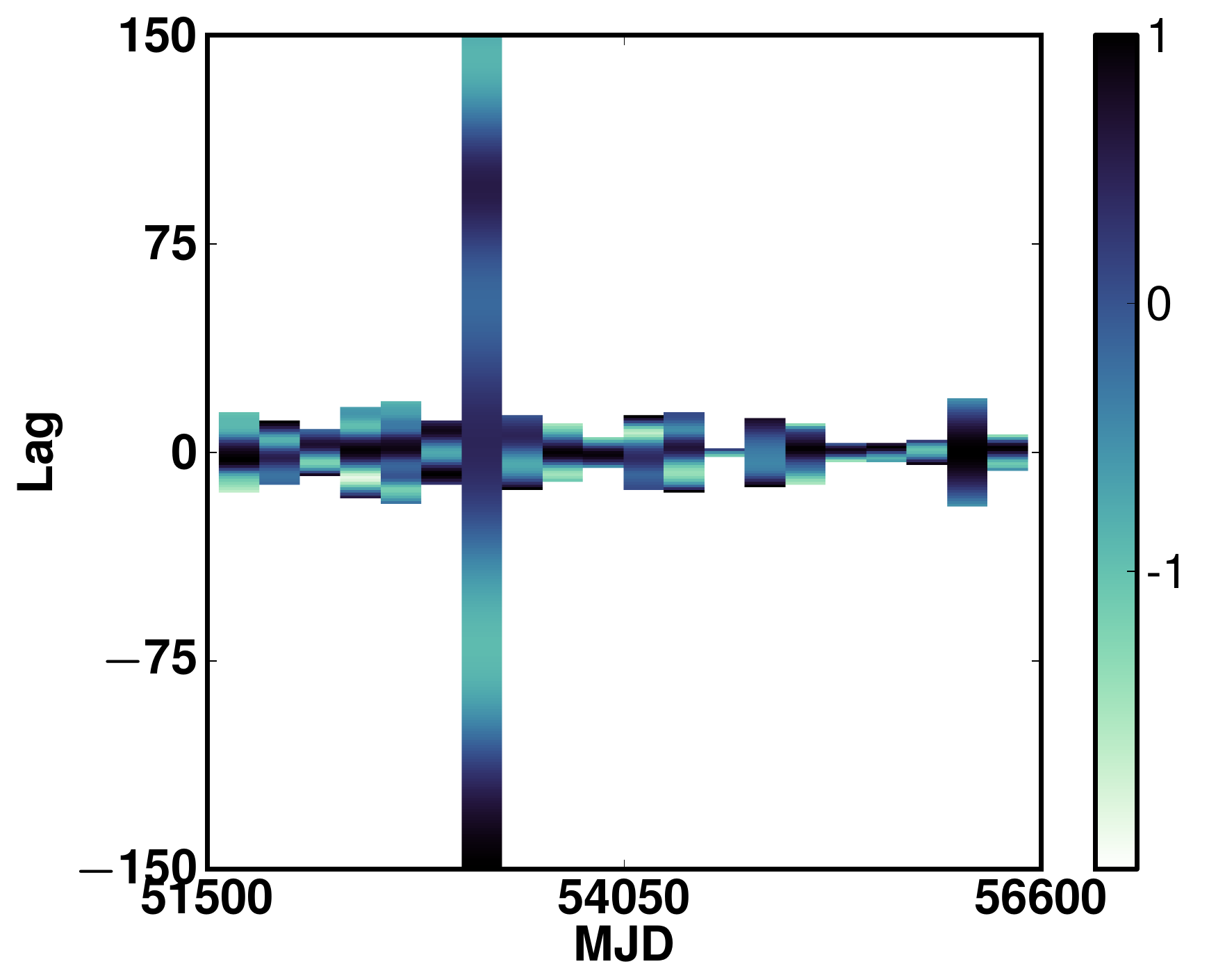}}\hfill
%  \caption{Time evolution of the CC time lags  for the  time periods of monitoring campaigns. 
%Note that  NGC 5548 exhibits clearly  symmetrical pattern  of behavior , while the largest values of CC time lags of NGC 4051  develops  over time almost %diagonally . No symmetry  or  diagonally distributed elements in the case of  Arp 102B, 3C 390.3 and E1821. }
  \label{fig:2}
\end{figure}

\begin{figure}[t!]
  \caption{Time evolution of the CC time lags  for the  time periods of monitoring campaigns. 
Note that  NGC 5548 exhibits clearly  symmetrical pattern  of behavior, while the largest values of CC time lags of NGC 4051  develops  over time almost diagonally. No symmetry  or  diagonally distributed elements  is seen in the case of  Arp 102B, 3C 390.3 and E1821+643.  Colorbars represent the  values of  Gaussian kernel based  CC coefficients. }
\end{figure}

    \begin{figure}[htb]
\captionsetup[subfloat]{farskip=0.1pt,captionskip=0.1pt}
\centering
   \subfloat[ Case 1 ]{%
    \includegraphics[width=7.5cm]{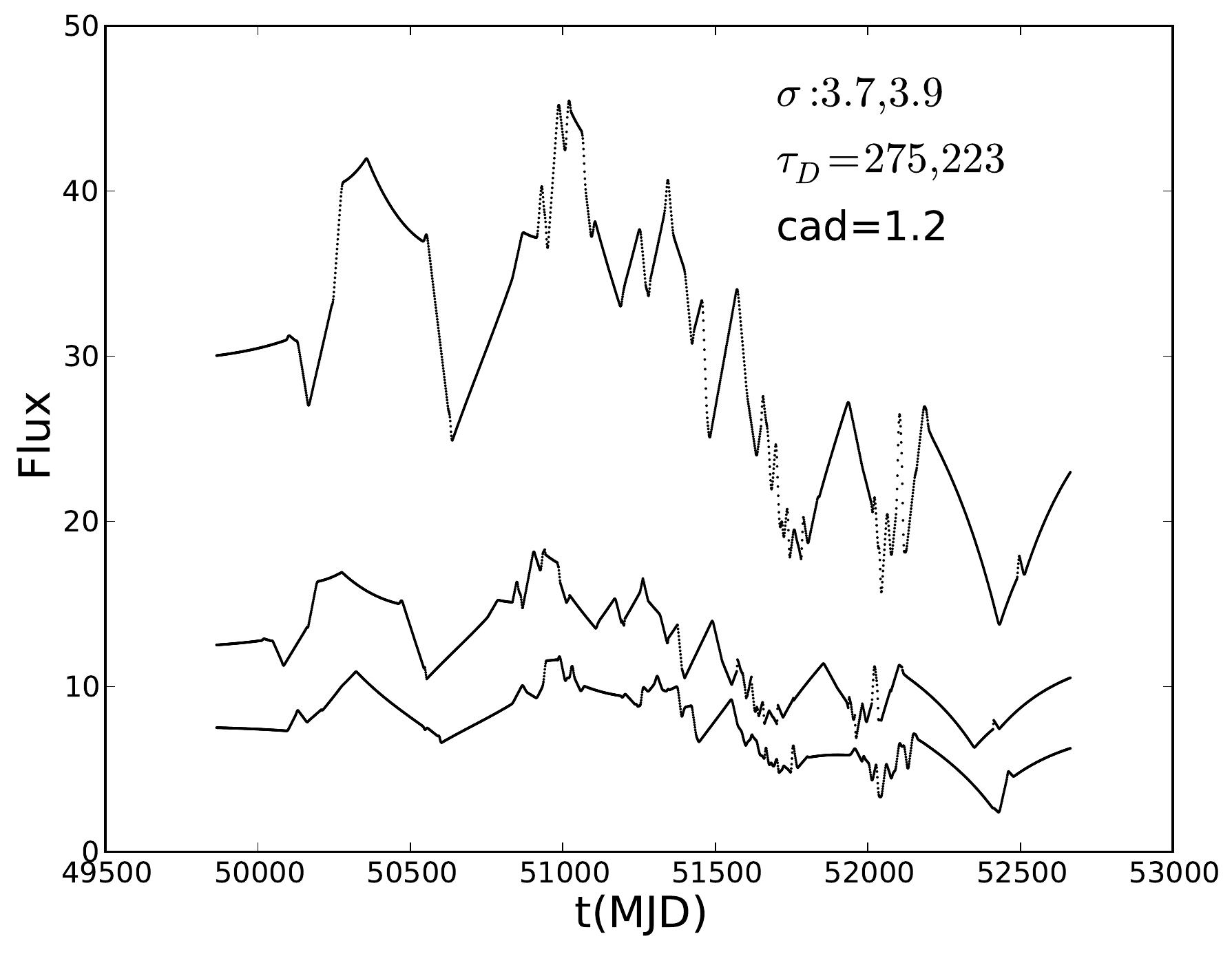}}\hfill

  \subfloat[ Case 2 ]{%
    \includegraphics[width=7.5cm]{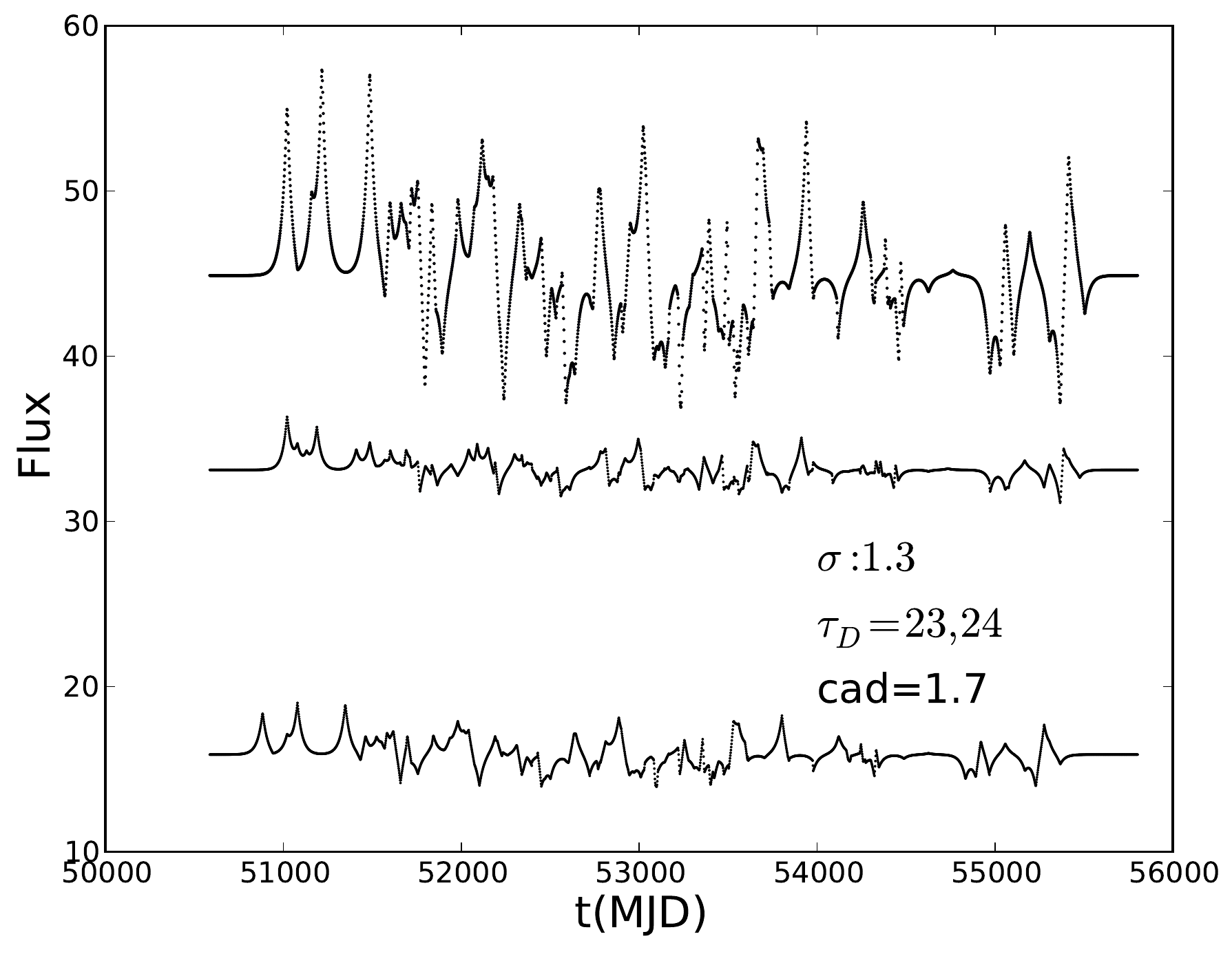}}\hfill\\

%      \subfloat[NGC 4051 ]{%
%    \includegraphics[width=4cm]{modelc4051.eps}}\hfill

   \subfloat[ Case 3]{%
    \includegraphics[width=7.5cm]{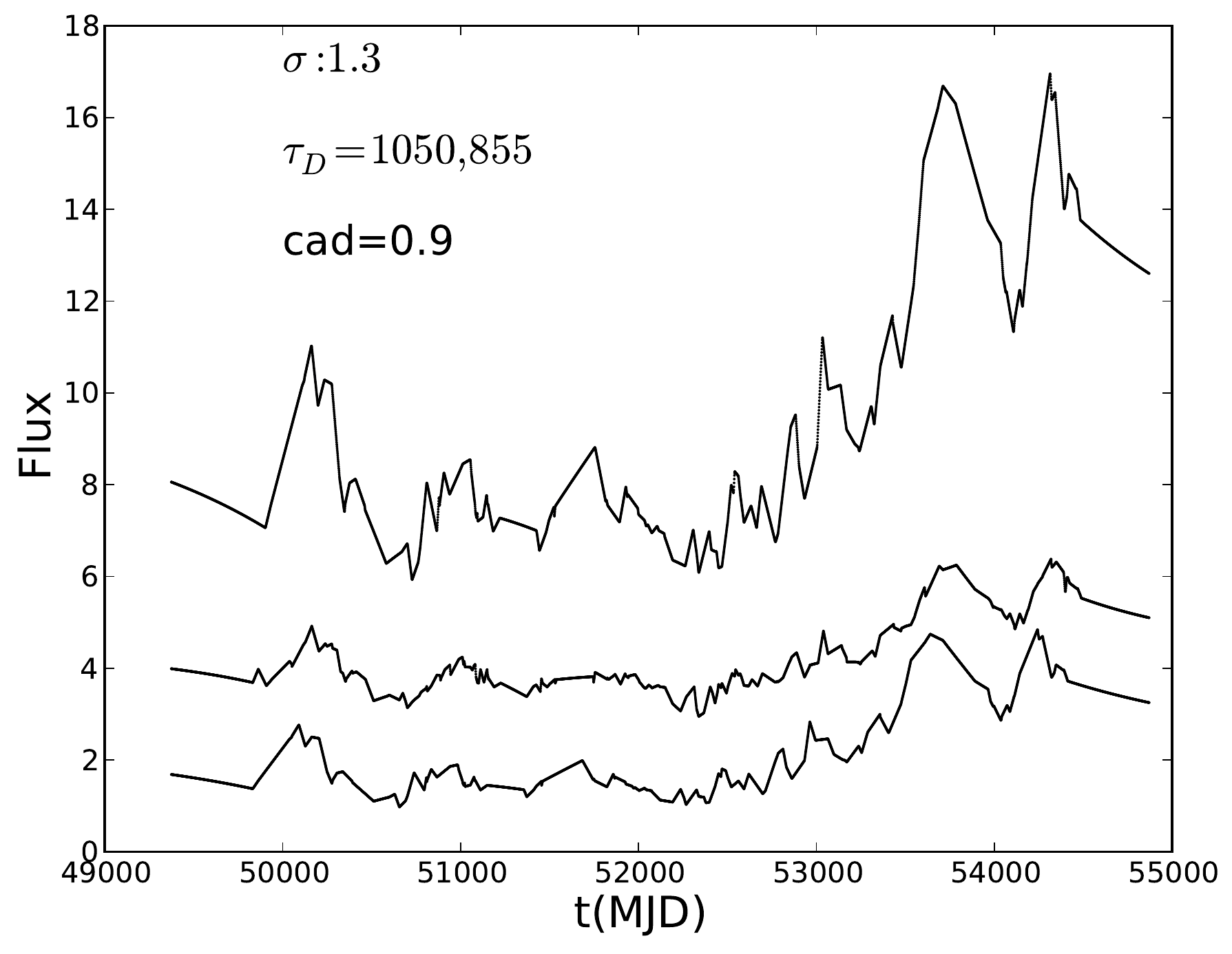}}\hfill \\

% \subfloat[ ]{%
 %   \includegraphics[width=4cm]{modelc1821.eps}}\hfill  
 \vspace*{-5mm} 
 \caption{ Synthetic light curves generated using  the variability parameters $\sigma$, $\tau_{D}$ and cadence, listed on  each panel, in units of mag, days, and days respectively.
  Assumed objects possessed  the variability parameters  of  NGC 5548 (Case 1, upper plot), Arp102B (Case 2, middle plot)  and 3C 390.3 (Case 3, bottom plot). 
  All  synthetic light curves cover original length of monitoring campaigns. On each panel the upper  and middle curve mimics emission lines while the bottom curve mimics
 continuum. }
  \label{fig:3}
  
  \end{figure}

     \begin{figure}[htb]

       \subfloat[Case 4 ]{%
    \includegraphics[width=8cm]{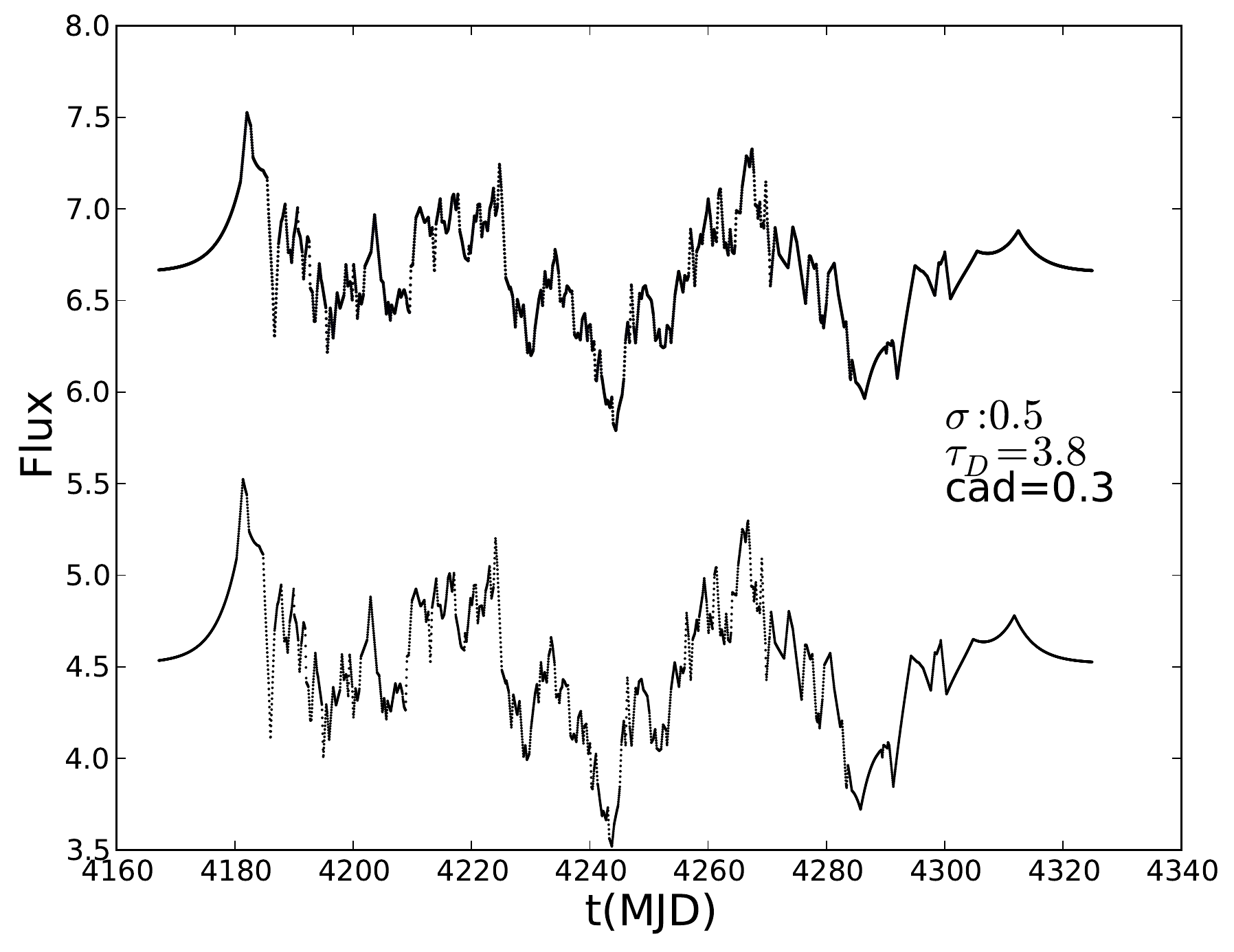}}\hfill

\subfloat[ Case 5 ]{%
  \includegraphics[width=8cm]{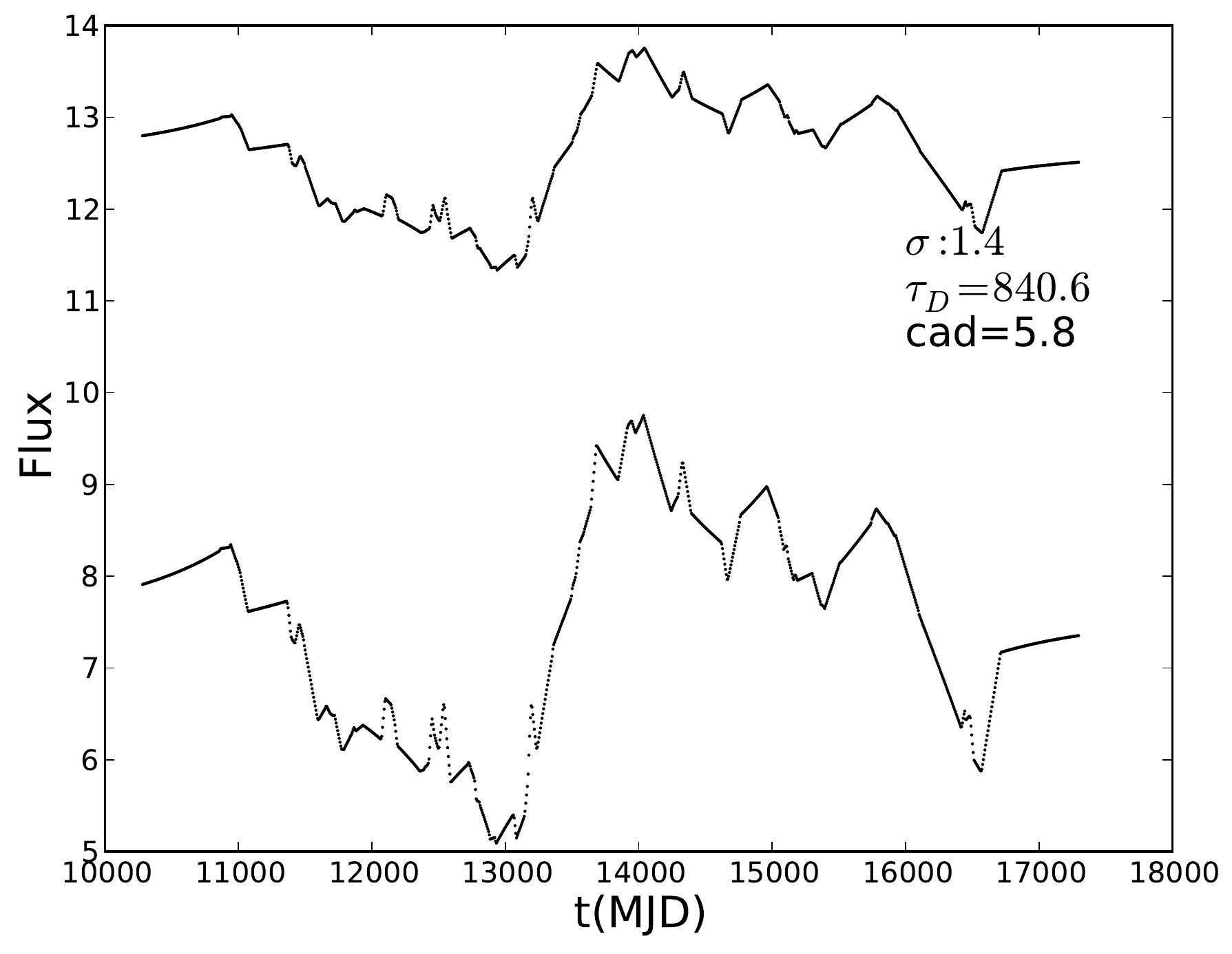}}\hfill  
 \caption{ The same s in Fig. 4  but for  the NGC 4051 (Case 4, top plot) and E1821+643 (Case 5, bottom plot).}
  \label{fig:4}
\end{figure}

 \begin{figure}[htb]
\captionsetup[subfloat]{farskip=1pt,captionskip=1pt}
\centering
  
  \subfloat[Case 1-UB ]{%
    \includegraphics[width=0.49\textwidth]{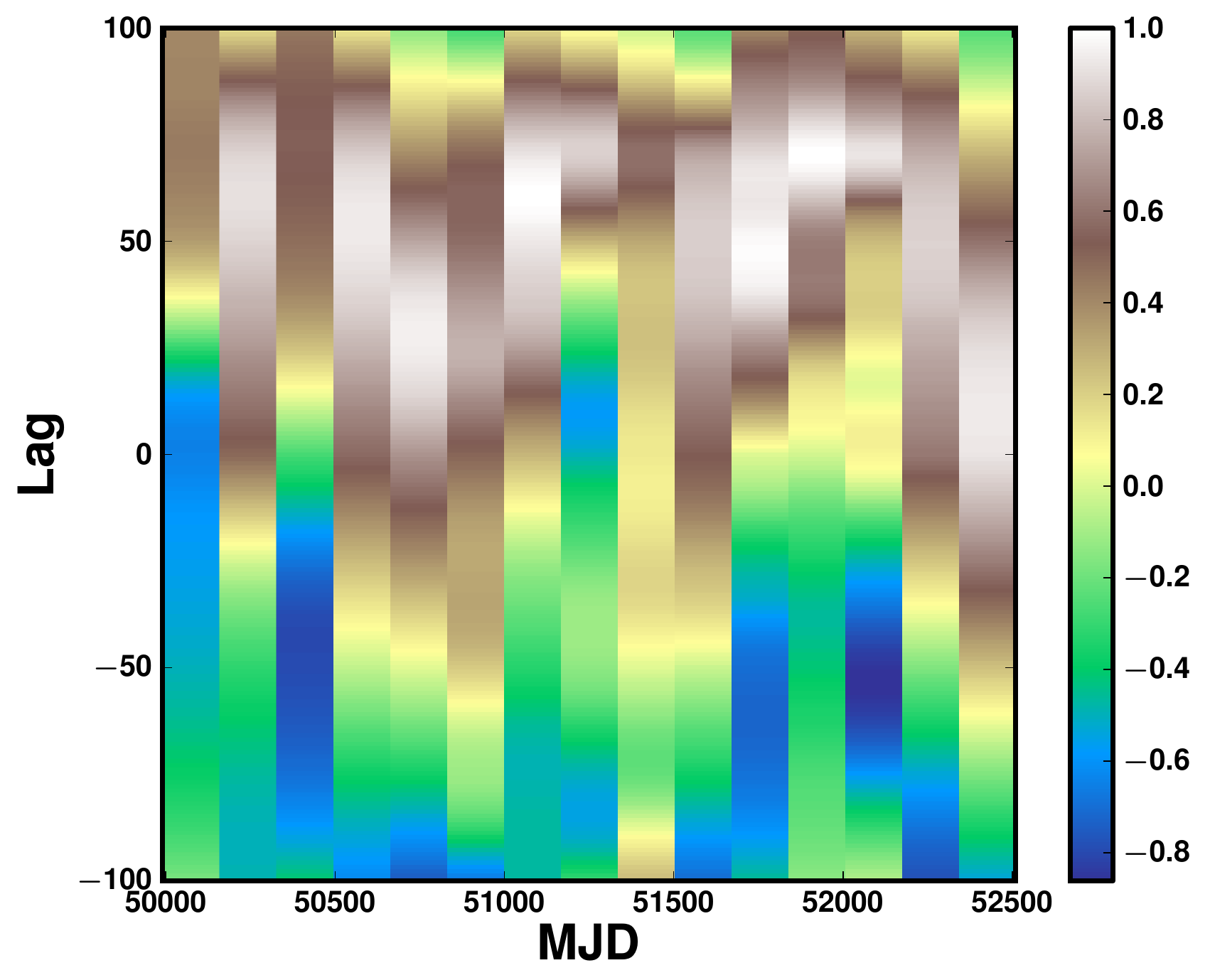}}\hfill
  \subfloat[ Case 1-MB ]{%
    \includegraphics[width=0.49\textwidth]{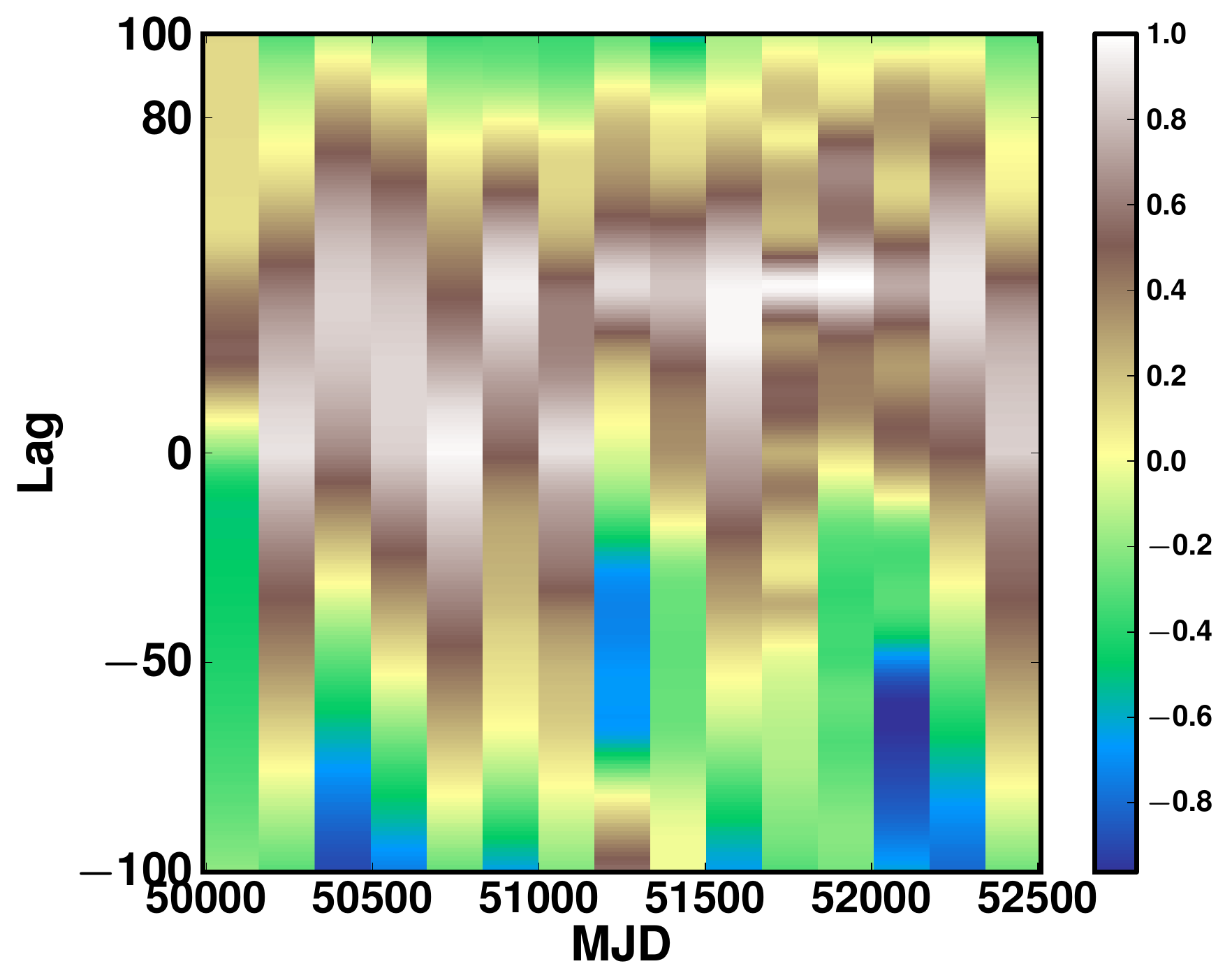}}\hfill\\

    \subfloat[Case 2-UB ]{%
    \includegraphics[width=0.49\textwidth]{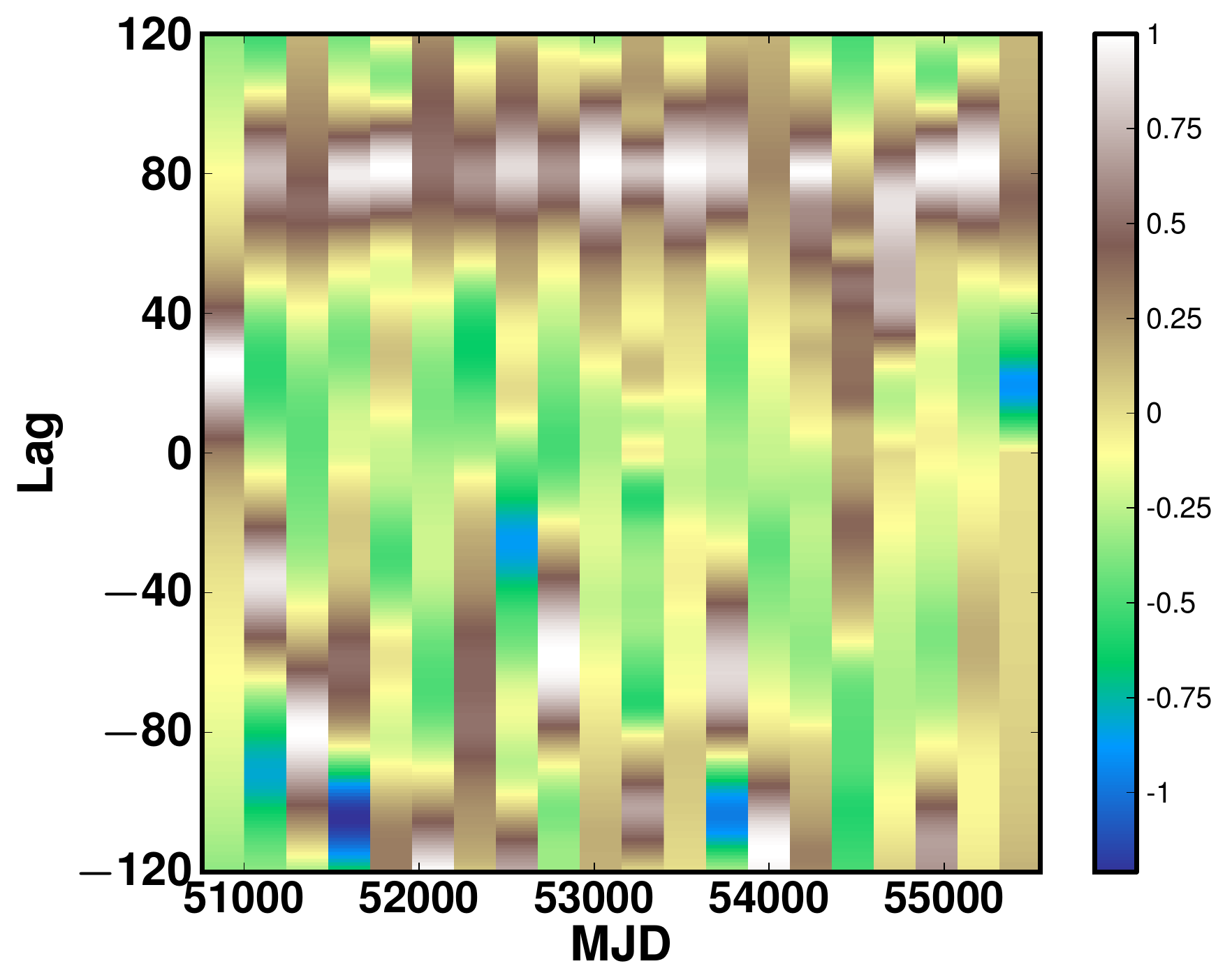}}\hfill
  \subfloat[ Case 2-MB ]{%
    \includegraphics[width=0.49\textwidth]{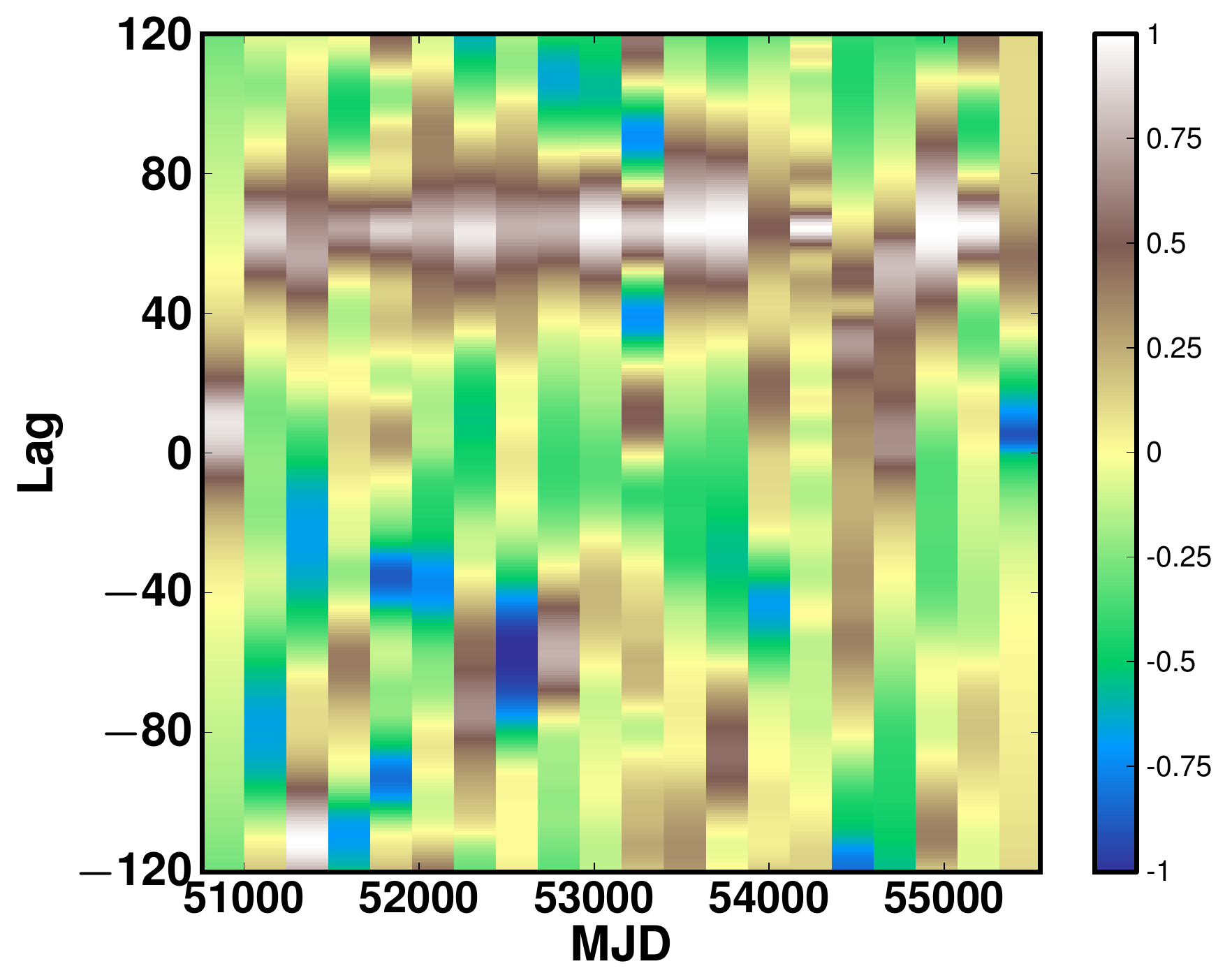}}\hfill \\

%      \subfloat[NGC 4051  H$\beta$   ]{%
%    \includegraphics[width=0.49\textwidth]{modelngc4051hb.eps}}\hfill\\

  \subfloat[  Case 3-UB ]{%
    \includegraphics[width=0.49\textwidth]{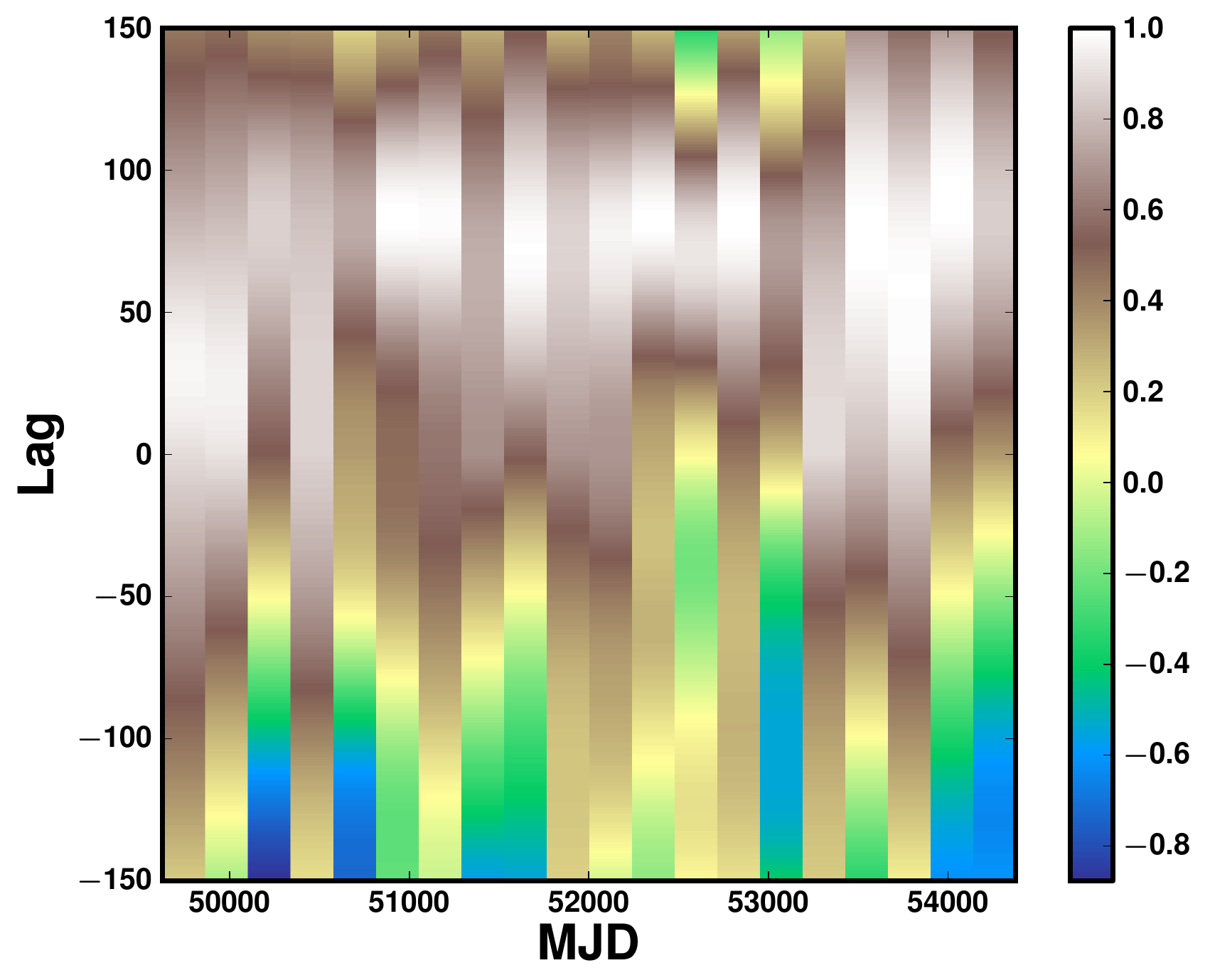}}\hfill
     \subfloat[Case 3-MB   ]{%
    \includegraphics[width=0.49\textwidth]{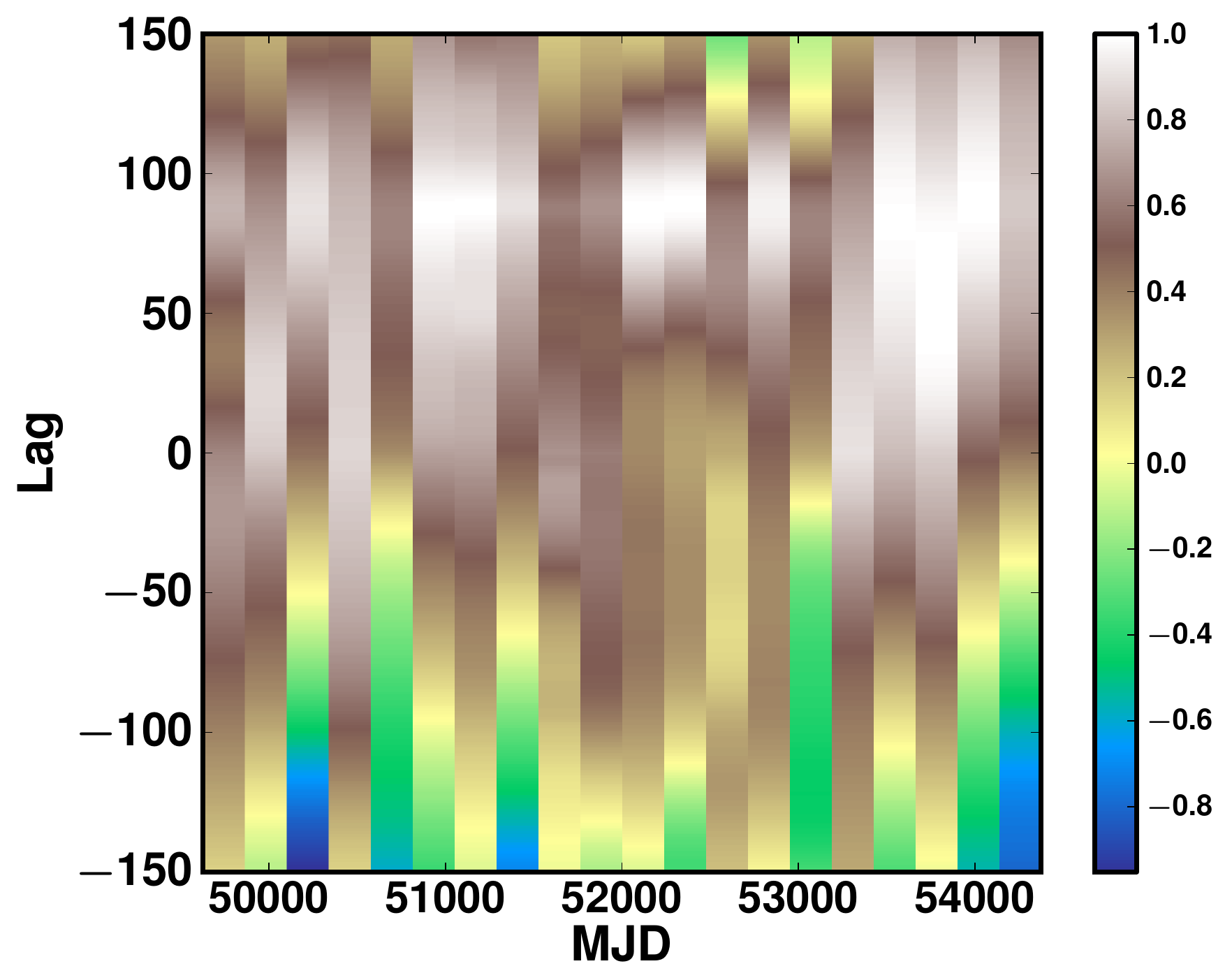}}\hfill\\
%  \subfloat[E1821+623 H$\beta$ ]{%
 %   \includegraphics[width=0.49\textwidth]{modele1821hb.eps}}\hfill
 \caption{ Gaussian kernel based method applied on  synthetic light curves  Cases 1, 2, 3 (see Fig. 3).  Left panels:  UB stands for the time evolution of  lags between 
the upper and bottom light curve from Fig.3. Right panels: MB stands for  time evolution of lags between middle  and bottom light curves from Fig. 3. }
  \label{fig:5}
\end{figure}

 \begin{figure}[htb]
\captionsetup[subfloat]{farskip=1pt,captionskip=1pt}
\centering

        \subfloat[Case 4 ]{%
   \includegraphics[width=0.49\textwidth]{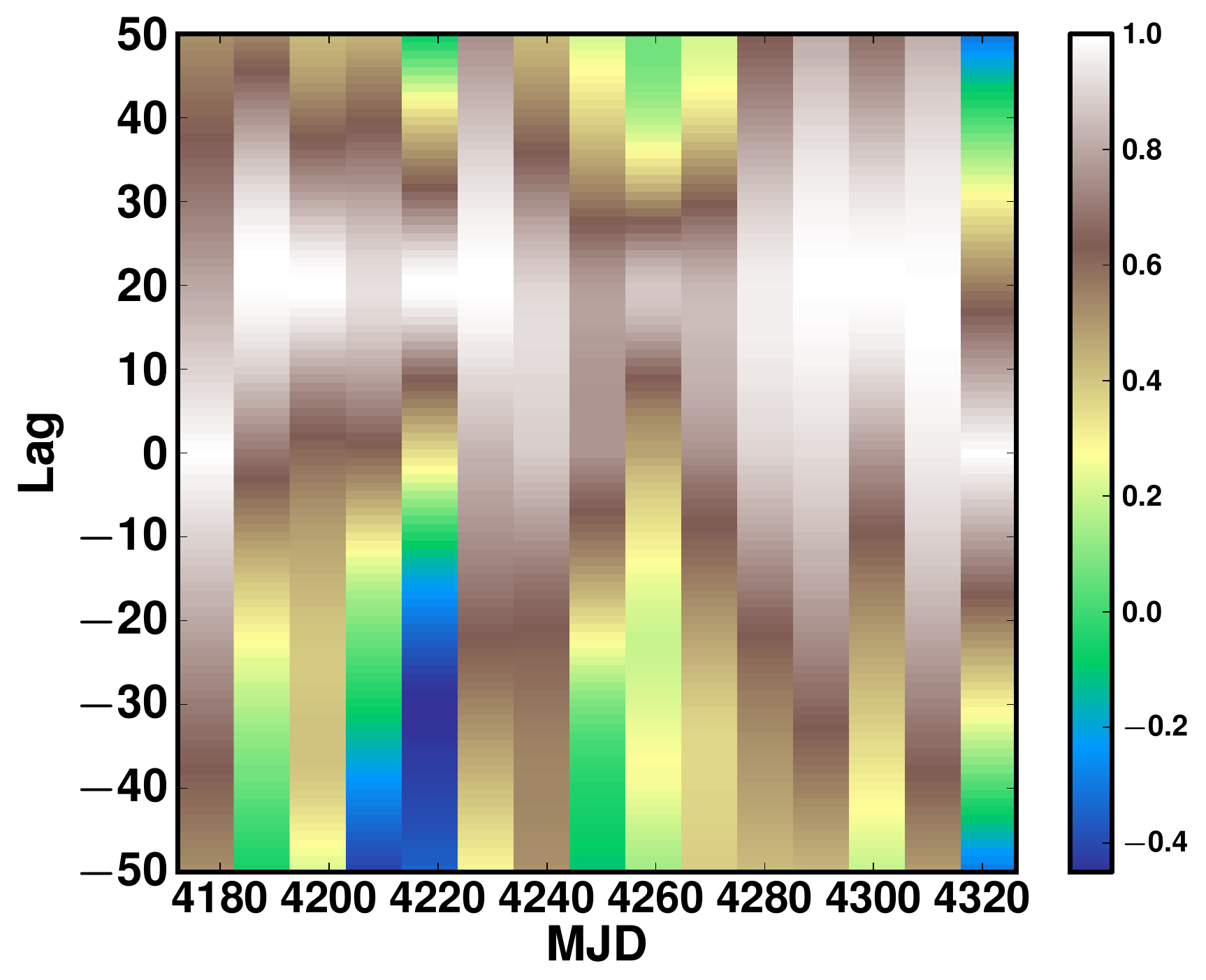}}\hfill\\

 \subfloat[Case 5 ]{%
   \includegraphics[width=0.49\textwidth]{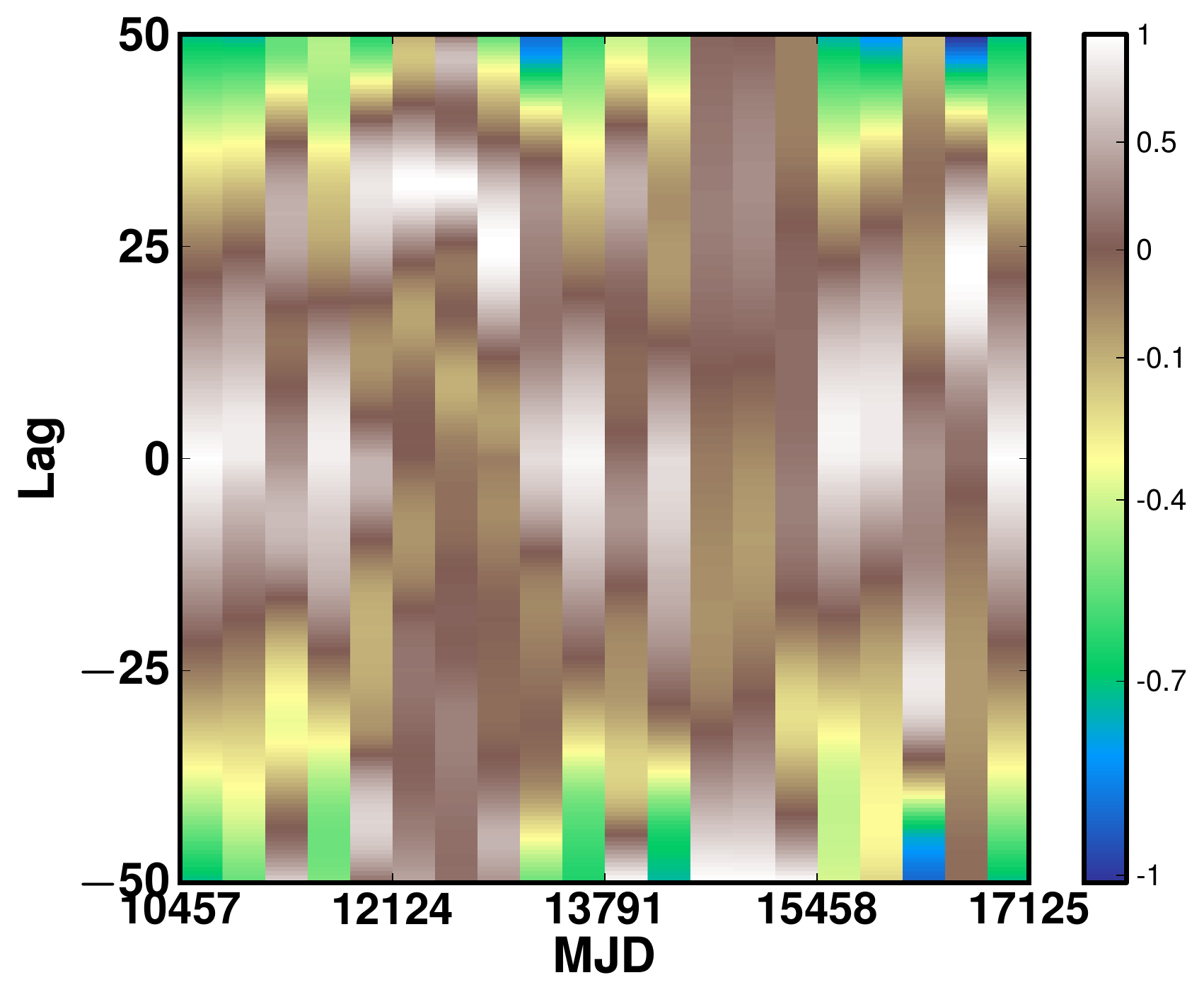}}\hfill
 \caption{ Same as  in Fig. 5 but for  Cases 4 and 5 (see  synthetic light curves given in the  Fig. 4).  }
  \label{fig:6}
\end{figure}

\begin{figure}[h]
%\sidecaption
% For example, with the graphicx style use
\includegraphics [scale=0.55]{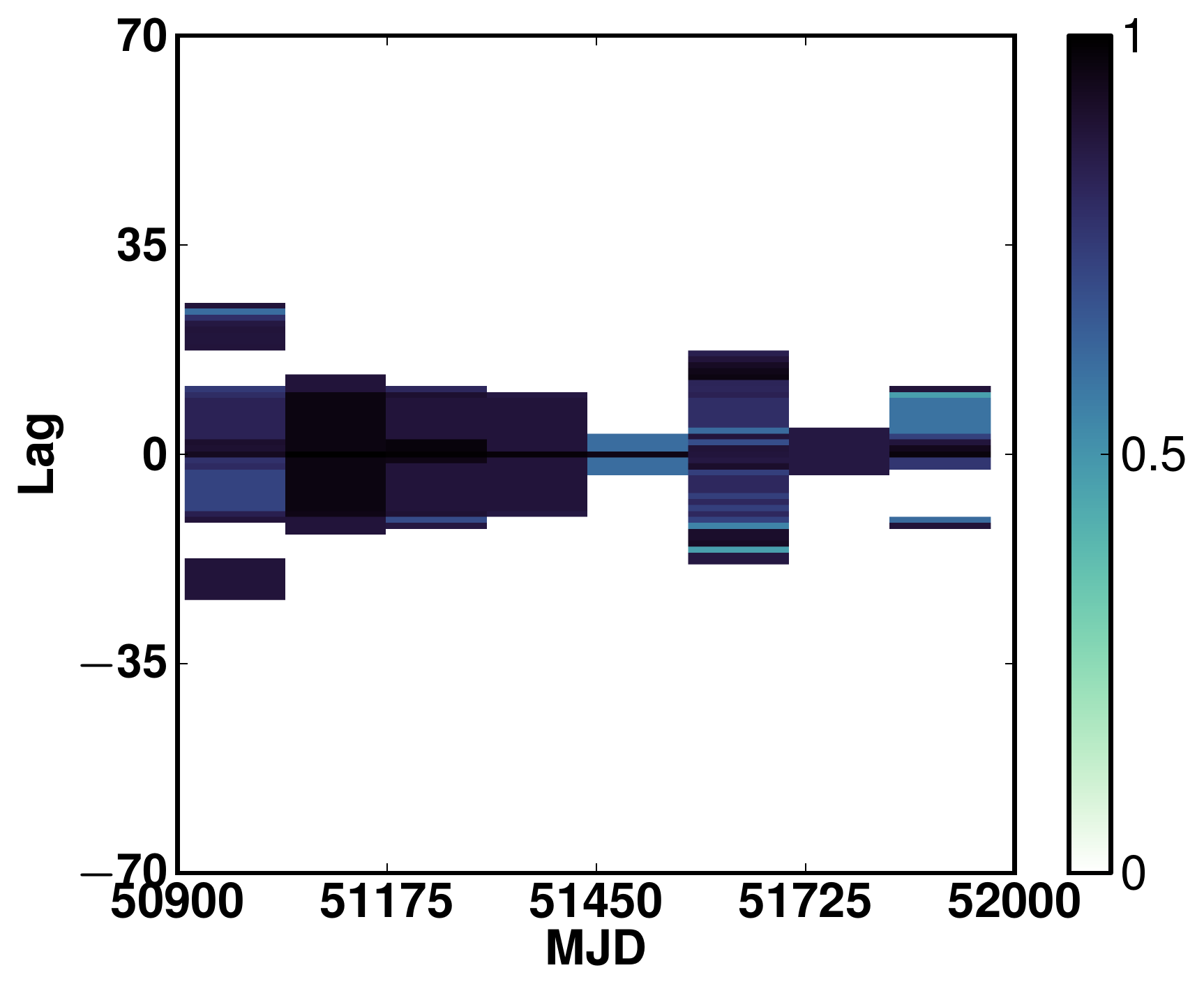}
%
% If no graphics program available, insert a blank space i.e. use
%\picplace{5cm}{2cm} % Give the correct figure height and width in cm
%
\caption{The  time evolution of MI based  time lag between  continuum and H$\beta$ line  of NGC 5548. }
\label{fig:7}       % Give a unique label
\end{figure}

\begin{table}
\begin{center}
%\resizebox{10cm}{!}{%
  \begin{tabular}{llll}
\hline
Object& Line &$\tau_{SPEAR}$  & $\tau_{ZDCF}$  \\
                    
\hline
 Arp 102B& H$\alpha$&${23.8}_{-18.8}^{+27.5}$& ${15.0}_{-13.8}^{+24.4}$  \\
         &           &                       &  ${157.1}_{-110.1}^{+20.3}$\\
 &H$\beta$&${47.6}_{-37.2}^{+57.2}$ & ${22.8}_{-20.9}^{+64.0} $ \\
 \hline
 3C 390.3& H$\beta$&$76.9_{-74.7}^{+79.1}$ &  $94.5_{-48.0}^{+27.1}$ \\ 
 \hline
 NGC 5548&H$\alpha$&$43.5_{-39.3}^{+47.7}$ & $27.0_{-5.7}^{+14.4}$  \\
 &H$\beta$&$45.4_{-43.0}^{+47.0}$  &  $49.2_{-7.7}^{+18.6}$ \\ 
 \hline
 NGC 4051&H$\beta$&$2.8_{-2.3}^{+3.1}$ & $2.6_{-1.1}^{+0.9}$ \\ 
 \hline
  
 E1821+643 &H$\beta$&$125.6_{-2.3}^{+3.4}$ & $26.3_{-24.1}^{+48.2}$ \\ 
              &                &                                         &    $137_{-19}^{+54}$              \\
 \hline

\end{tabular}
%}

\end{center}

\caption{The time lags determined by  ZDCF and SPEAR method (see Kova{\v c}evi{\'c} et al. (2014)).}
\label{table3}
\end{table}

\begin{table}
\begin{center}
%\resizebox{10cm}{!}{%
  \begin{tabular}{llll}
\hline
Object& $ \check{ \tau}$  & $\sigma^{+}$& $\sigma^{-}$   \\
                    
\hline
 Arp 102B& 14.4& 39.1 & 27.4 \\

 \hline
 3C 390.3& 51.4 & 77.1& 76.9 \\ 
 \hline
 NGC 5548&10.8&47.5 &48  \\
 
 \hline
 NGC 4051&2.2& 8.1&9.1\\ 
 \hline
 
\end{tabular}
%}

\end{center}
\caption{The weighted mean   of    time lags ($ \check{ \tau}$)  obtained by other authors (compiled in Kova{\v c}evi{\'c} et al. (2014)) and their combined errors
 ($\sigma^{+}$,  $\sigma^{-}$),  calculated by method of Barlow (2003).}
\label{table4}
\end{table}

\section{Conclusion}

To conclude, the present work is a  continuation  of  our previous publication  (Kova{\v c}evi{\'c} et al. (2014)) on  time series analysis of    enlarged data set:  Arp 102B, 3C 390.3, NGC 5548,  NGC 4051 and in addition E1821+643. Here we examined  both non-stationarity of  their time series by means of ADF and    time evolution of  time lags  between continuum and line curves by means of Gaussian kernel based CC and MI. 
We found strong evidence for non-stationarity of all time series in  ADF test regime without constant terms, which corresponds to modeling damped random walk. 

 The application of Gaussian kernel based CC through observed data set shows specific temporal behavior of time lags between continuum and lines. Running  this  method to surrogate  light curves (which are stationary and having constant lags) the  results  still show strong variation in lags (excluding the case of cloned curves of  Arp 102B).
In those surrogate cases with significantly variable time lags,  the certain discrepancies between the DRW time scales  and used sliding time windows of Gaussian kernel based CC are noticed.
So  the recorded time lag variations in the  cases of  NGC 5548, 3C 390.3, NGC 4051 and E1821+643 can be a consequence  of instability of the method due to above mentioned  anomalies.  From the other hand, it seems that  DRW time scale of Arp 102B has 'Goldilocks' value  so the method shows more stability (time lags of its cloned curves do not vary with time notably) and detected variations could  correspond to the real behavior of  time lag over observed time period.
 Nevertheless, tracking of MI time evolution of time lags has failed due to a small number of points in used  time windows.
The results  highlight specific issues so the further analysis is necessary.

\begin{acknowledgement}
This work was supported by the Ministry of Education , Science and Technological Development  of Republic of Serbia through 
the project Astrophysical Spectroscopy of Extragalactic Objects (176001) and RFBR (grants N12-02-01237a, 12-02-00857a) (Russia) 
and CONACYT research grants 39560, 54480, and 151494 (Mexico)..
\end{acknowledgement}
%

%%%%%%%%%%%%%%%%%%%%%%%% referenc.tex %%%%%%%%%%%%%%%%%%%%%%%%%%%%%%
% sample references
% %
% Use this file as a template for your own input.
%
%%%%%%%%%%%%%%%%%%%%%%%% Springer-Verlag %%%%%%%%%%%%%%%%%%%%%%%%%%
%
% BibTeX users please use
% \bibliographystyle{}
% \bibliography{}
%
%\biblstarthook{References may be \textit{cited} in the text either by number (preferred) or by author/year.\footnote{Make sure that all references from the list are cited in the text. Those not cited should be moved to a separate \textit{Further Reading} section or chapter.} The reference list should ideally be \textit{sorted} in alphabetical order -- even if reference numbers are used for the their citation in the text. If there are several works by the same author, the following order should be used: 

\end{document}